\documentclass[11pt]{article}
\usepackage{amsmath, amsfonts, setspace, url, hyperref, verbatim, xcolor, cite, graphicx, todonotes, cancel, slashed, mathtools}
\usepackage[T1]{fontenc}
\usepackage[utf8]{inputenc}
\usepackage{fullpage}
\usepackage{tikz}
\usetikzlibrary{arrows,positioning,fit,shapes,snakes,backgrounds,tikzmark}
\usepackage{pgfplots}
\usetikzlibrary{shapes.geometric,calc}

\newcommand{\be}{\begin{equation}}
\newcommand{\ee}{\end{equation}}
\numberwithin{equation}{section}

\newcommand{\cH}{\mathcal{H}}

\newcommand{\Gfour}{\mathrm{SL}(5)}

\newcommand{\Gsix}{E_{6(6)}}

\newcommand{\Hfour}{\mathrm{SO}(5)}

\definecolor{vub}{RGB}{0,52,154}
\definecolor{vubo}{RGB}{255,102,0}
\definecolor{redd}{RGB}{255,40,40}
\definecolor{r}{RGB}{228,32,20}
\definecolor{o}{RGB}{238,69,4}
\definecolor{y}{RGB}{253,228,1}
\definecolor{g}{RGB}{108,160,0}
\definecolor{b}{RGB}{0,162,203}
\definecolor{i}{RGB}{120,42,117}

\usepackage[utf8]{inputenc}
\usepackage[framemethod=tikz]{mdframed}
\usetikzlibrary{calc}
\usepackage{chngcntr}

\counterwithin{ex}{section}

\makeatletter
\def\mdf@@mynote{}
\define@key{mdf}{mynote}{\def\mdf@@mynote{#1}}

\mdfdefinestyle{mystate}{
skipabove={1\baselineskip},
linewidth=0.5pt,
innertopmargin=1\baselineskip,
roundcorner=10pt,
backgroundcolor=black!0,
linecolor=black,
singleextra={
  \node[xshift=10pt,draw=black,fill=white,rounded corners,anchor=west] at (P-|O) %
  {\strut{\itshape  }\ifdefempty{\mdf@@mynote}{}{\itshape\bfseries \mdf@@mynote}};
},
firstextra={
  \node[xshift=10pt,draw=black,fill=white,rounded corners,anchor=west] at (P-|O) %
  {\strut{\itshape  }\ifdefempty{\mdf@@mynote}{}{\itshape\bfseries \mdf@@mynote}};
}
}

\mdfdefinestyle{mystateb}{
skipabove={1\baselineskip},
linewidth=1pt,
innertopmargin=1\baselineskip,
roundcorner=10pt,
backgroundcolor=white!5,
linecolor=vub!50,
singleextra={
  \node[xshift=10pt,thick,draw=vub!50,fill=b!0,rounded corners,anchor=west] at (P-|O) %
  {\strut{\itshape  }\ifdefempty{\mdf@@mynote}{}{\bf\mdf@@mynote}};
},
firstextra={
  \node[xshift=10pt,thick,draw=vub!50,fill=b!0,rounded corners,anchor=west] at (P-|O) %
  {\strut{\itshape  }\ifdefempty{\mdf@@mynote}{}{\bf\mdf@@mynote}};
}
}

\mdfdefinestyle{mystater}{
skipabove={1\baselineskip},
linewidth=1pt,
innertopmargin=1\baselineskip,
roundcorner=10pt,
backgroundcolor=white!5,
linecolor=r!50,
singleextra={
  \node[xshift=10pt,thick,draw=r!50,fill=b!0,rounded corners,anchor=west] at (P-|O) %
  {\strut{\itshape  }\ifdefempty{\mdf@@mynote}{}{\bf\mdf@@mynote}};
},
firstextra={
  \node[xshift=10pt,thick,draw=r!50,fill=b!0,rounded corners,anchor=west] at (P-|O) %
  {\strut{\itshape  }\ifdefempty{\mdf@@mynote}{}{\bf\mdf@@mynote}};
}
}

\mdfdefinestyle{mystateg}{
skipabove={1\baselineskip},
linewidth=1pt,
innertopmargin=1\baselineskip,
roundcorner=10pt,
backgroundcolor=white!5,
linecolor=g!50,
singleextra={
  \node[xshift=10pt,thick,draw=g!50,fill=g!0,rounded corners,anchor=west] at (P-|O) %
  {\strut{\itshape  }\ifdefempty{\mdf@@mynote}{}{\bf\mdf@@mynote}};
},
firstextra={
  \node[xshift=10pt,thick,draw=g!50,fill=g!0,rounded corners,anchor=west] at (P-|O) %
  {\strut{\itshape  }\ifdefempty{\mdf@@mynote}{}{\bf\mdf@@mynote}};
}
}

\newmdenv[style=mystate,nobreak=true]{state}
\newmdenv[style=mystater,nobreak=true]{stater}
\newmdenv[style=mystateg,nobreak=true]{stateg}
\newmdenv[style=mystateb,nobreak=true]{stateb}

\makeatother

\makeatother

\newcommand{\fM}{\mathcal{M}}
\newcommand{\fN}{\mathcal{N}}
\newcommand{\fP}{\mathcal{P}}
\newcommand{\fQ}{\mathcal{Q}}
\newcommand{\fK}{\mathcal{K}}
\newcommand{\fL}{\mathcal{L}}
\newcommand{\fA}{\mathcal{A}}
\newcommand{\fB}{\mathcal{B}}
\newcommand{\fC}{\mathcal{C}}
\newcommand{\fD}{\mathcal{D}}
\newcommand{\fE}{\mathcal{E}}
\newcommand{\fF}{\mathcal{F}}

\newcommand{\ui}{\underline{i}}
\newcommand{\uj}{\underline{j}}
\newcommand{\uk}{\underline{k}}

\newcommand{\EDA}{Exceptional Drinfeld Algebra}

\begin{document}

\begin{titlepage}

\vfill

\begin{center}
	\baselineskip=16pt  
	
	{\Large \bf  \it Exploring Exceptional Drinfeld Geometries}
	\vskip 1cm
	{\large \bf  Chris D. A. Blair$^{a,}$\footnote{\tt cblair@vub.ac.be}, Daniel C. Thompson$^{a,b,}$\footnote{\tt d.c.thompson@swansea.ac.uk}, Sofia Zhidkova$^{a,}$\footnote{\tt Sofia.Zhidkova@vub.be}}
	\vskip .6cm
	{\it  
			$^a$ Theoretische Natuurkunde, Vrije Universiteit Brussel, and the International Solvay Institutes, \\ Pleinlaan 2, B-1050 Brussels, Belgium \\ \ \\
			$^b$ Department of Physics, Swansea University, \\ Swansea SA2 8PP, United Kingdom \\ \ \\}
	\vskip 2cm
\end{center}

\begin{abstract} 
We explore geometries that give rise to a novel algebraic structure, the Exceptional Drinfeld Algebra, which has recently been proposed as an approach to study generalised U-dualities, similar to the non-Abelian and Poisson-Lie generalisations of T-duality.  
This algebra is generically not a Lie algebra but a Leibniz algebra, and can be realised in exceptional generalised geometry or exceptional field theory through a set of frame fields giving a generalised parallelisation.  
We provide examples including ``three-algebra geometries'', which encode the structure constants for three-algebras and in some cases give novel uplifts for $CSO(p,q,r)$ gaugings of seven-dimensional maximal supergravity.
We also discuss the M-theoretic embedding of both non-Abelian and Poisson-Lie T-duality.
\end{abstract}

\vfill

\setcounter{footnote}{0}
\end{titlepage}
\tableofcontents 
 
 
\vspace{1em}\noindent 

\section{Introduction}
 
The textbook T-duality symmetry of string theory that applies in backgrounds with Abelian isometries is a cornerstone of the duality web that ultimately leads to M-theory \cite{Hull:1994ys,Witten:1995ex}.   
Less standard is the application of T-duality to backgrounds whose isometry group is non-Abelian \cite{delaOssa:1992vci}.  
While its status as a precise duality in either $\alpha^\prime$ and $g_s$ expansions is not fully resolved, at the very least non-Abelian T-duality (NATD) is a useful tool as a solution generating symmetry of Type II supergravity (for a review see \cite{Thompson:2019ipl}).  More exotic still are applications of T-duality to backgrounds which have no isometries at all.   Poisson-Lie (PL) T-duality, introduced by Klim\v{c}\'ik and Severa \cite{Klimcik:1995dy,Klimcik:1995ux}, provides situations where such a non-isometric duality can be realised. This is made possible when the target spaces have a certain   Poisson-Lie symmetry property giving rise to  an unexpectedly rich algebraic structure encoded by a Drinfeld double, $\mathfrak{d}$ \cite{Drinfeld:1986in}.\footnote{The Drinfeld double $\mathfrak{d}$ is an even-dimensional Lie algebra that can be decomposed into two sub-algebras $\mathfrak{d} = \mathfrak{g} +\tilde{\mathfrak{g}}$ that are maximally isotropic with respect to an ad-invariant inner product of split signature.   The Jacobi identity of $\mathfrak{d}$ enforces a cocycle compatibility condition between $\mathfrak{g}$ and $\tilde{\mathfrak{g}}$.} Despite this lack of isometry, the corresponding non-linear sigma models can actually exhibit classical (and quantum) integrability   \cite{Klimcik:2008eq}. Close connections between integrability and Poisson-Lie duality have come under renewed focus with holographic motivation following the development of the integrable $\eta$   \cite{Klimcik:2008eq} and related $\lambda$  \cite{Sfetsos:2013wia}  deformations applied to the $AdS_5 \times S^5$ superstring in \cite{Delduc:2013qra} and \cite{Hollowood:2014qma} respectively. 

Poisson-Lie geometries (i.e. those for which PL T-duality can be realised) can at first sight seem convoluted, especially when presented in terms of the regular geometric data consisting of the metric and Kalb-Ramond two-form.  However, when viewed using generalised geometry the situation is radically improved; the PL property of the target space is encapsulated \cite{Hassler:2017yza} by a generalised parallelisation \cite{Grana:2008yw,Lee:2014mla}.   This consists of a set of generalised frame fields that span the generalised tangent bundle, $TM\oplus T^\star M$, and which furnish the Drinfeld double algebra  under the generalised Lie derivative.   Moreover there is a natural candidate for the extended target space that appears in both the world-sheet doubled sigma-model \cite{Klimcik:1996nq,Hull:2009sg}  and in the Double Field Theory approach\cite{Hassler:2017yza,Demulder:2018lmj}, namely the group $D=\exp \mathfrak{d}$.\footnote{The discussion here is adapted to the case where the physical target space  $M$ is a group manifold $M = G \cong D/ \tilde{G}$ with $G=\exp \frak{g}$ and $\widetilde{G} =\exp \tilde{\frak{g}}$.  However,  when $M$ can be constructed as a double coset, $M = H \backslash D / \widetilde{G}$, similar ideas apply both from the world-sheet \cite{Klimcik:1996np} and target space \cite{Demulder:2019vvh} perspectives.}

The U-duality symmetry of M-theory can also be viewed as a generalisation of T-duality, arising when one combines the perturbative T-duality symmetry with non-perturbative S-dualities.  Until recently, there has been no hint of whether U-duality admits non-Abelian or generalised versions.  A proposal for the algebraic structure that would underlie such dualities has been introduced in \cite{Sakatani:2019zrs,Malek:2019xrf} and called the \EDA{} (EDA).

Roughly an EDA is an algebra $\mathfrak{d}_n$,  defined by a bracket, $[\bullet, \bullet]: \mathfrak{d}_n \otimes \mathfrak{d}_n \to \mathfrak{d}_n$, which is need not be antisymmetric but obeys the Leibniz identity, and which admits a Lie subalgebra $\mathfrak{g}$, of dimensions $n$ or $n-1$.  Moreover $\mathfrak{g}$ can be considered a maximally isotropic subalgebra in a sense we shall make more precise later. For the case of $n\leq 4$, that shall be our concern here, the data of an EDA can be interpreted as consisting of a Lie-algebra $\mathfrak{g}$ together with a three-algebra $\tilde{\mathfrak{g}}$ that are restricted to obey a cocycle compatibility condition.  A key point of \cite{Sakatani:2019zrs,Malek:2019xrf} was that the EDA can be realised by a  generalised Leibniz parallelisation for the exceptional tangent bundle $TG \oplus \wedge^2 T^\star G$ thus echoing the set up of Poisson-Lie T-duality and allowing this framework to be used to generate solutions using the ideas of generalised Scherk-Schwarz reductions.
Some features of the geometry, and the membrane interpretation, were then given in \cite{Sakatani:2020iad}, while a classification of all possible EDAs for the case of $n=3$ was made in \cite{Hlavaty:2020pfj}.

In this paper, we shall explore the geometry associated to this new M-theoretic algebraic structure in a number of explicit examples.  These examples reveal intriguing connections to several topics. We study geometries which encode the structure constants of three-algebras, which naturally show up amongst the structure constants of the \EDA.  Here we can also connect with a class of $CSO$ gaugings of 7-dimensional maximal supergravity. Hence, we get for free out of our construction some simple new uplifts for these gaugings. These uplifts could be regarded as ``non-Abelian U-duals'', in some sense, of spheres with flux.  We will also describe the embedding of Poisson-Lie T-duality into this set-up in some detail, revealing a construction whereby the \EDA{} involves augmenting the Drinfeld double with a spinor representation.

Our presentation will make frequent usage of some technical results within Exceptional Field Theory which, to allow for completeness but avoid distraction, have been included as appendix material here. (For a detailed review, see \cite{Berman:2020tqn}.)

\section{The $\Gfour$ \EDA}

\subsection{The algebra}

We begin by specifying the \EDA{} in the case of the group $E_{4(4)} = \Gfour$.
We introduce five-dimensional fundamental $\Gfour$ indices $\fA,\fB=1\,\dots,5$.
The generators of the \EDA{} live in the ten-dimensional antisymmetric representation, and we can label these with a pair of antisymmetric five-dimensional indices, $T_{\fA \fB} = - T_{\fB \fA}$.
The brackets of the generators are
\be
 [ T_{\fA \fB} , T_{\fC \fD} ] = \frac{1}{2}   F_{\fA \fB , \fC \fD}{}^{\fE \fF} T_{\fE \fF} \,,
\label{algebra}
\ee
(where the factor of $1/2$ is inserted to avoid overcounting)
and these need not be antisymmetric. We do require the Leibniz identity
\be
\big[ T_{\fB \fB^\prime} , [ T_{\fC \fC^\prime}, T_{\fD \fD^\prime}  ] \big]
= \big[  [ T_{\fB \fB^\prime} ,  T_{\fC \fC^\prime}] , T_{\fD \fD^\prime}  \big] 
+  \big[   T_{\fC \fC^\prime}, [ T_{\fB \fB^\prime} , T_{\fD \fD^\prime}]  \big] \,,
\label{leibnizid1}
\ee
which in terms of the structure constants leads to
\be
\frac{1}{2} F_{\fB \fB^\prime , \fE \fE^\prime}{}^{\fA \fA^\prime} F_{\fC \fC^\prime , \fD \fD^\prime}{}^{\fE \fE^\prime} 
- 
\frac{1}{2} F_{\fC \fC^\prime , \fE \fE^\prime}{}^{\fA \fA^\prime} F_{\fB \fB^\prime , \fD \fD^\prime}{}^{\fE \fE^\prime} 
= 
\frac{1}{2} F_{\fB \fB^\prime , \fC \fC^\prime}{}^{\fE \fE^\prime} F_{\fE \fE^\prime , \fD \fD^\prime}{}^{\fA \fA^\prime} 
\label{qc10}
\,.
\ee
If the bracket is antisymmetric, this reduces to the usual Jacobi identity. 

More generally, the constraint \eqref{qc10} is the same as the quadratic constraint of gauged supergravity.
This link -- or equivalently the fact that we are restricting to Leibniz algebras which can arise from a generalised parallelisation of $\Gfour$ exceptional geometry -- also motivates the assumption that the structure constants can be decomposed into irreducible representations as 
\begin{equation}
    \begin{aligned}
        F_{\fA \fB , \fC \fD}{}^{\fE \fF} &=& 4 F_{\fA \fB [\fC}{}^{[\fE} \delta_{\fD]}^{\fF]} \,,\quad  
        F_{\fA \fB \fC}{}^{\fD} &=& Z_{\fA \fB \fC}{}^{\fD} + \frac{1}{2} \delta^{\fD}_{[\fA} S_{\fB] \fC} - \frac{1}{6} \tau_{\fA \fB} \delta^{\fD}_{\fC} - \frac{1}{3} \delta^{\fD}_{[\fA} \tau_{\fB]\fC} \, , 
    \end{aligned}
\end{equation}
 where $\tau_{\fA\fB} = - \tau_{\fB\fA}$, $S_{\fA\fB} = S_{\fB \fA}$ and $Z_{\fA\fB\fC}{}^{\fD} = Z_{[\fA \fB\fC]}{}^{\fD}$, $Z_{\fA\fB \fC}{}^{\fC} = 0$. 
This means that the only $\Gfour$ irreducible representations appearing in the structure constants of our Leibniz algebra are those specified by the linear constraint of gauged maximal supergravity in seven-dimensions \cite{Samtleben:2005bp}.

Now we impose the further conditions that make this $\Gfour$ Leibniz algebra into an \EDA{}.
We require that there is a  Lie subalgebra $\mathfrak{g}\subset \mathfrak{d}_4$ which is isotropic in the sense that\footnote{Note that a systematic construction of generalised frames corresponding to a given set of generalised fluxes was set out in \cite{Inverso:2017lrz} in which a similar condition plays a necessary role: it really just ensures that the section condition of Exceptional Field Theory is satisfied.}
\be
\epsilon^{\fA \fB \fC \fD \fE}  T_{\fA \fB} \otimes T_{\fC \fD} \Big|_{\mathfrak{g}} = 0 \,,
\label{isotropic}
\ee
and we further require this isotropic to be \emph{maximal} in the sense that appending any extra generator to $\mathfrak{g}$ will violate \eqref{isotropic}.
This means that it will have either dimension 4 or 3, and so can be interpreted (borrowing terminology from Exceptional Field Theory) as the {\em physical} subalgebra in either an M-theory or type IIB background, respectively.  To  articulate this condition in a more invariant fashion we can say that alongside $\mathfrak{d}_n$ we must specify a ``pure spinor'' $\Lambda$ in an appropriate representation\footnote{In DFT this would actually be a spinor representation, in ExFT it is not generically spinorial but will obey a purity constraint projecting out certain representations in the tensor product of $\Lambda$ with itself.} of $E_{n(n)}$  which acts linearly on the $\mathfrak{d}_n$ vector space schematically as $\Lambda \bullet T$. We then demand that the kernel of  this action,  $\mathfrak{g}=  \ker(\Lambda)  $  be a Lie subalgebra.  There are different choices for $\Lambda$ that will result in a subalgebra $\mathfrak{g}$ of dimension $n$, which we call an M-theory section, and dimension $n-1$ which we shall call a IIB-theory section.  This pure spinor approach is essentially the same as that used to define solutions to the so-called section condition of Exceptional Field Theory \cite{Berman:2012vc, CederwallKorea}.

For the case of $\Gfour$,   in the IIB-theory  section the pure spinor $\Lambda$ is in the $\overline{\mathbf{10}}$ and the purity condition is that $\Lambda^{[\fA \fB} \Lambda^{ \fC \fD]}= 0$.  The linear action is defined by 
 $$
  \Lambda  \bullet T : = \Lambda^{\fA \fC} T_{\fC \fB} - \frac{1}{5}  \Lambda^{\fC \fD} T_{\fC \fD} \delta^{\fA}{}_{\fB}.  
 $$
 As an example consider $\Lambda^{45} = - \Lambda^{54} =1$ with the other components zero.  Evidently this is pure and it is such that it defines 
\begin{equation}
 \ker(\Lambda)  = \textrm{span} \{T_{12}, T_{13}, T_{23} \}\, . 
\end{equation}
  In the M-theory section the pure spinor $\Lambda$ is in the $\mathbf{5}$, the purity constraint is automatic and no further conditions are placed on $\Lambda$.  The action on generators is 
\begin{equation}
 \Lambda  \bullet T : = \Lambda_{[\fA }T_{\fB \fC]} \, .
\end{equation}
  Consider taking $\Lambda_\fA = \delta_{\fA,5}$, in which case 
\begin{equation} 
 \ker(\Lambda) = \textrm{span} \{ T_{a 5} | a = 1 \dots 4   \} \, . 
\end{equation}
 We will continue now in this M-theory section, and decompose indices  as   $\fA=(a,5)$, where $a=1,\dots,4$ such that   the physical subalgebra is generated by the generators $t_a \equiv T_{a5}$, with Lie algebra structure constants $f_{ab}{}^c$.

In terms of the irreducible representations, the \EDA{} is wholly defined in terms of the Lie algebra structure constants $f_{ab}{}^c$ along with $S_{ab}, \tau_{ab}$ and $\tau_{a5}$, with:
\begin{equation}
\begin{aligned}
S_{55} = 0 \,,\quad
Z_{abc}{}^5 = 0 \,,\quad
Z_{ab5}{}^5 = \frac{2}{3} \tau_{ab}\,,\quad
Z_{abc}{}^d = -\tau_{[ab} \delta_{c]}^d \, ,
\\
S_{a5} = - \frac{2}{3} \tau_{a5} - \frac{4}{3} f_{ab}{}^b \,,\quad
Z_{ab5}{}^c = - f_{ab}{}^c - \frac{2}{3} \delta_{[a}^c f_{b]d}{}^d \,.
\label{edaflux2}
\end{aligned} 
\end{equation}
To write down the algebra explicitly, we combine $S_{ab}$ and $\tau_{ab}$ into a ``dual'' structure constant with three upper antisymmetric indices given by 
\be
\tilde f^{abc}{}_d =   \frac{1}{4}  \epsilon^{abce} ( S_{de} + 2 \tau_{de} ) \,.
\label{fStau}
\ee  
If we further define the ``dual'' generators $\tilde t^{ab} \equiv \frac{1}{2} \epsilon^{abcd} T_{cd}$, then the \EDA{} can then be written as 
\be
\begin{split}
[t_a, t_b ] & =  f_{ab}{}^c t_c \,,  \\
[t_a, \tilde t^{bc} ] &  =    2  f_{ad}{}^{[b} \tilde t^{c]d} - \tilde f^{bcd}{}_a t_d - \frac{1}{3} \frak{L}_a  \tilde t^{bc}   \, ,   \\ 
 [\tilde t^{bc} , t_a ] & =     3 f_{[de}{}^{[b} \delta_{a]}^{c]} \tilde t^{de} 
  + \tilde  f^{bcd}{}_a t_d 
  +  \frak{L}_d  \delta_a^{[b} \tilde t^{cd]} 
   \,, \\
[\tilde t^{ab}, \tilde t^{cd} ] & =   2 \tilde f^{ab[c}{}_e \tilde t^{d]e}   \, ,
\end{split}
\label{EDAf}
\ee
in which we introduced  the combination $\frak{L}_a =\tau_{a5} - f_{ad}{}^d $. With $\frak{L}_a = 0$   this presentation closely resembles    the structure of a Drinfeld double.  However crucially   this bracket has a symmetric part that vanishes if and only if  
 \be 
  \frac{2}{3}  \frak{L}_{[d}\delta^{c}_{e]} +   f_{de}{}^{c}=0  \, , \quad 
 \tau_{ab} = 0   \,.
 \ee
In addition to the Jacobi identity on $\mathfrak{g}$,  the Leibniz closure conditions \eqref{qc10} enforce that the dual structure constants obey the fundamental identity of a three-algebra
\begin{equation}
\tilde{f}^{a b g}{}_c \tilde{f}^{def}{}_g - 3 \tilde{f}^{g [de}{}_c \tilde{f}^{f] a b}{}_g = 0 \, . \label{THREEEEE} 
\end{equation} 
There are also a set of compatibility equations between $\tilde{f}^{abc}{}_d$ and $f_{ab}{}^c$ which include in particular a   condition 
\begin{equation}
 6f_{f [a}{}^{[c}\tilde{f}^{de]f}{}_{b]} +f_{ab}{}^f \tilde{f}^{cde}{}_f + \frac{2}{3}  \tilde{f}^{cde}{}_{[a}\frak{L}_{b]} = 0 . \label{COCY} 
\end{equation} 
When $\frak{L}_a =0$ this last condition states that the dual structure constants, viewed as a map  $\tilde{f}: \mathfrak{g} \to \wedge^3 \frak{g}$ define a  $\wedge^3 \frak{g}^*$ valued one-cochain.  

\subsection{The generalised geometry realisation}

A geometric realisation of this algebra can be achieved using as data the left-invariant forms $l^a$ and dual vector fields $v_a$, obeying $\iota_{v_a} l^b = \delta_a^b$, of a group manifold $G$, together with a trivector $\lambda^{abc}$ and a scalar $\alpha$ that are required to  obey differential conditions:
\be
d l^a = \frac{1}{2} f_{bc}{}^a l^b \wedge l^c \,,\quad L_{v_a} v_b=-f_{ab}{}^c v_c \,, 
\label{dgroup}
\ee
\be
d \lambda^{abc} = \tilde f^{abc}{}_d l^d + 3 f_{ed}{}^{[a} \lambda^{bc]d} l^e + \frac{1}{3} \lambda^{abc} \frak{L}_d l^d \,,
\label{dlambda}
\ee
\be
L_{v_a}\ln \alpha  =\frac{1}{3} \frak{L}_a \equiv  \frac{1}{3} ( \tau_{a5} - f_{af}{}^f )\,.
\label{dalpha} 
\ee
Below, we will often write the trivector $\lambda^{abc}$ in its dualised form 
\be
\lambda^{abc} = \epsilon^{abcd} \lambda_d \,,\quad
\lambda_a = \frac{1}{3!} \epsilon_{bcda} \lambda^{bcd} \,.
\ee
These data can be naturally understood in terms of a generalised frame field using $\Gfour$ exceptional generalised geometry or $\Gfour$ exceptional field theory \cite{Berman:2010is, Berman:2011cg, Coimbra:2011ky, Coimbra:2012af, Hohm:2013pua}.
We provide the necessary background material in appendix \ref{howtoexft}, and will only summarise the key details here.
A generalised frame is a section of the generalised tangent bundle $TM \oplus \Lambda^2 T^*M$, where $M$ denotes a four-dimensional manifold, and so we can write $E_{\fA \fB} = ( e_{\fA\fB}, \omega_{(2) \fA \fB})$ in terms of vector field $e_{\fA \fB}$ and a two-form $\omega_{(2) \fA \fB}$.  Under  the generalised Lie derivative (for more see appendix \ref{gld}) which acts as
\be
\mathcal{L}_{E_{\fA\fB}} E_{\fC \fD} = ( L_{e_{\fA\fB}} e_{\fC \fD} , L_{e_{\fA\fB}} \omega_{(2)\fC \fD} - \iota_{e_{\fC\fD}} d \omega_{(2) \fA \fB} )\,,
\ee
the frames are constructed such that they obey  
\be
 \mathcal{L}_{ E_{\fA \fB}}  E_{\fC \fD}
= - \frac{1}{2} F_{\fA \fB ,\,\fC \fD}{}^{\fE \fF} E_{\fE \fF}\, ,  
 \label{10alg}
\ee
where in general the quantities $F_{\fA \fB ,\,\fC \fD}{}^{\fE \fF}$ give non-constant ``generalised fluxes'' defined as in appendix \ref{howtoexft}. 
We are interested in the case where a set of frames can be found with constant fluxes, in which case their generalised Lie derivatives \eqref{10alg} furnish a geometric realisation of a Leibniz algebra.

We can achieve such a realisation of our \EDA.
First, we decompose our 10-dimensional generalised frame as
\be
E_a \equiv E_{a 5} \,,\quad E^{ab} \equiv \frac{1}{2} \epsilon^{abcd} E_{cd}  
\,,
\ee
and specify that, in terms of pairs of vectors and two-forms, these are given by
\be
E_a =(  v_a , 0 )  \,,\quad E^{ab} = ( \lambda^{abc} v_c, \alpha l^a \wedge l^b ) \,.
\label{10frame}
\ee
The differential conditions \eqref{dgroup}, \eqref{dlambda} and \eqref{dalpha} ensure that the algebra of frames \eqref{10alg} reproduces the \EDA{} \eqref{EDAf} subject to the imposition of some algebraic constraints which take the form:
\be
\begin{split}
0   = f_{[ab}{}^d \lambda_{c]} + 6 \lambda_{[a} \frak{L}_b   \delta_{c]}^d \,,\quad 
0   = \tau_{[ab} \lambda_{c]} \,.
\end{split} 
\label{adjinv} 
\ee
These constraints ensure that the structure constants of the EDA are invariant under an adjoint action of $G= \exp\frak{g}$ \cite{Sakatani:2019zrs,Malek:2019xrf}. They are also what is needed to ensure that the structure constants are indeed constant.

In what follows, it will be convenient to package the same data into a frame field $\tilde{E}_{\fA}$ in the $\bf{5}$ representation i.e. as sections of the bundle 
$(\mathbb{R} \oplus \Lambda^3 T^*M ) \otimes ( \det T^*M )^{-3/10}$.
Here the weight factor is such that the frame has unit determinant when viewed as a five-by-five matrix (see appendix \ref{howtoexft} for more details).
This matrix is given by 
\begin{equation}\label{eq:5frameunitdet}
    \tilde{E}^{\fM}{}_{\fA}=  \Delta^{-\frac{1}{2}}  \begin{pmatrix}
      l^\frac{1}{2} \alpha^\frac{1}{2}  v^i{}_a & 0   \\
          l^{-\frac{1}{2}} \alpha^{-\frac{1}{2}} \lambda_a & l^{-\frac{1}{2}} \alpha^{-\frac{1}{2}}   
    \end{pmatrix} \, ,
\end{equation}
where $l\equiv \det l^a{}_i$ and $\Delta = \alpha^\frac{3}{5} l^\frac{1}{5}$ is a corrective weight whose interpretation in terms of the determinant of the external 7-dimensional metric is explained in appendix \ref{howtoexft}.

\subsection{The geometry} 
In the $E_{n(n)}$ Exceptional Generalised Geometry (EGG) / Exceptional Field Theory (ExFT) approach to supergravity an artificial splitting is made into $n$ internal directions (coordinates of which we denote $x$) and $D=11-n$ external directions (coordinates of which we denote $X$).  This splitting allows the field content\footnote{More precisely the bosonic field content is packaged into representations of $E_{n(n)}$ while the fermions (which play no role in the discussion here) form representations of the maximal compact subgroup.}  of the supergravity to be reassembled into appropriate representations of the $E_{n(n)}$.    

In the case at hand, $n=4$,  the degrees of freedom associated to the ``internal''  four-dimensional metric, $g_{ij}$,  and three-form, $C_{ijk}$, parametrise the coset $\Gfour/\Hfour$. This coset can be described using a generalised frame or equivalently a $\Hfour$-invariant matrix $m_{\fM \fN}$ called the generalised metric. 
The technical details of how to extract the conventional geometric data from a generalised metric are presented in the appendix. In particular note that we have one extra piece of geometric data, namely the scalar $\Delta \equiv \Delta(x)$ (or equivalently $\alpha$), which is related to the determinant of the external metric. 

Here we will consider generalised metrics admitting a particular factorised form using the generalised frame field \eqref{eq:5frameunitdet}, such that
\be
m_{\fM \fN} ( X,x) = \tilde E^{\fA}{}_{\fM} (x) \tilde E^{\fB}{}_{\fN} (x) \bar m_{\fA \fB} (X) \,,
\label{SS}
\ee
where $\bar m_{\fA\fB}(X)$ denotes an $\Gfour/\Hfour$ coset element depending only on the external coordinates $X$.
This factorised form of eq.~\eqref{SS} is known as a generalised Scherk-Schwarz reduction ansatz.
It is now well-established that, starting with EGG/ExFT, such an ansatz
gives rise to lower-dimensional maximal gauged supergravities \cite{Berman:2012uy,Hohm:2014qga} (this idea was pioneered in the half-maximal case in DFT in \cite{Aldazabal:2011nj, Geissbuhler:2011mx, Grana:2012rr}). The structure constants of the \EDA{}  are interpreted as the embedding tensor which specifies the gauging of this theory,   and the matrix  $\bar m_{\fA \fB}$ contains the scalars of the gauged supergravity. 

One can regard two separate generalised frames $E^{\fA}$ and $E^\prime{}^{\fA}$ producing the same \EDA, up to some $\Gfour$ transformation acting on the indices $\fA$, but possibly depending on different choices of the physical coordinates,  as being generalised U-dual in the sense that they will both reduce to the same 7-dimensional theory.   

A key point here is that to complete the geometries given by the EDA frame fields as fully-fledged solutions of 11-dimensional supergravity one needs to determine the external sector by solving the equations of the resulting lower dimensional gauged supergravity.  Conversely, given a solution of the gauged supergravity whose embedding tensor matches the form of an EDA, then the ansatz \eqref{SS} provides an uplift.  
Our immediate aim however is not to construct full supergravity solutions, instead we wish simply to gain some intuition for the sort of geometries that arise when the generalised frame fields of the EDA are used to construct the internal metric.  To this end let us simply set $\bar m_{\fA \fB}(X) = \delta_{\fA \fB}$ and set to zero off-diagonal components of fields i.e.  those with mixed four-dimensional and seven-dimensional indices.  Using the dictionary reproduced in full in appendix \ref{dictionary}, we can, as in \cite{Sakatani:2020iad}, work out the geometry giving rise to the Exceptional Drinfeld Algebra 
\be
\begin{split}
ds_{11}^2 & =\alpha^{2/3} (1+\lambda_c \lambda^c )^{1/3} \left( ds_7^2
+ \frac{1}{1+\lambda_c \lambda^c}   (\delta_{ab} + \lambda_a \lambda_b )  l^a \otimes l^b  
\right) \\
&= \alpha^{2/3} (1+\lambda_c \lambda^c )^{1/3} ds_7^2 +ds_4^2   \,,\\
C_{(3)} & = -\frac{1}{6} \frac{\alpha}{1+\lambda_c \lambda^c} \lambda_{bcd}   l^b \wedge l^c \wedge l^d  \,,
\label{generalbg}
\end{split}
\ee
where we use $\delta_{ab}$ to contract Lie algebra indices.
%
%

\section{Three-algebra geometries}

We will start by exploring geometries with
\be
f_{ab}{}^c = 0 \,,\quad \tilde f^{abc}{}_d \neq 0 \,,
\ee
which we shall refer to as three-algebra geometries.  The analogue of such cases in terms of non-Abelian T-duality would be the geometries that one obtains \emph{after} dualising from a geometry with a group manifold symmetry, $f_{ab}{}^c \neq 0$, $\tilde f^{ab}{}_c = 0$. 

The corresponding \EDA{} is most transparently expressed in terms of the undualised generators 
\be
\begin{split} 
[T_{a5}, T_{b5} ] & = 0 \,,\\
[ T_{a5}, T_{bc} ] & =  \frac{1}{2} ( S_{a[b} + 2 \tau_{a[b} ) T_{c]5} = - [T_{bc}, T_{a5}]\,,\\
[ T_{ab}, T_{cd} ] & =  - \tau_{ab} T_{cd} + ( S_{c][b} + 2 \tau_{c][b} )T_{a][d} \,.
\end{split}
\ee
When $\tau_{ab} = 0$, this is the Lie algebra $CSO(p,q,r+1)$, $p+q+r=4$, as is clear from diagonalising $S_{ab}$ such that $S_{\mathcal{A}\mathcal{B}} \sim \text{diag} ( \underbrace{+1 , \dots + 1 }_{p}, \underbrace{-1,\dots,-1}_q,
\underbrace{0,\dots,0}_{r+1})$.
When $\tau_{ab} \neq 0$ we have a genuine Leibniz algebra.
The conditions for closure are 
\be
S_{a[b} \tau_{cd]} = 0 \,,\quad \tau_{[ab} \tau_{cd]} = 0 \,,
\ee
which are also what are required for the final equation of \eqref{adjinv} to hold. 
The only solutions can be organised according to the rank of $S_{ab}$ assuming the latter has been diagonalised:\footnote{If $S_{ab}$ is not diagonal then the constraints on $\tau_{ab}$ will be different, as will the form of the algebra, but this will be related by a similarity transform.}
\begin{itemize}
\item $S_{ab}$ has rank 4 or 3, then $\tau_{ab} = 0$,
\item $S_{ab}$ has rank 2, say $S_{11} \neq 0$, $S_{22} \neq 0$, then we can have $\tau_{12} \neq 0$,
\item $S_{ab}$ has rank 1, say $S_{11} \neq 0$, then we can have $\tau_{12}, \tau_{13}, \tau_{14} \neq 0$,
\item $S_{ab}$ has rank 0, then we can have either $\tau_{12}, \tau_{13}, \tau_{14} \neq 0$ or $\tau_{12}, \tau_{13}, \tau_{23} \neq 0$ (or other choices related by relabellings of the indices).
\end{itemize} 
 
In order to realise this algebra using a generalised frame, we introduce 4-dimensional coordinates $x^i$ and take
\be
l^a{}_i = \delta^a_i \,,\quad \lambda^{abc} = \tilde f^{abc}{}_d x^d \,,\quad \alpha = \text{constant} \,,
\ee
(where $x^a \equiv \delta^a_i x^i$). To extract the geometry, we note that
\be
\lambda_a = \frac{1}{6} \epsilon_{bcda} \tilde f^{bcd}{}_e x^e =  \frac{1}{4} ( S_{ab} - 2 \tau_{ab} ) x^b \,,
\ee
which we can use in the general formulae \eqref{generalbg}.

If we choose the coordinates $x^i$ to be periodic, then this corresponds to a U-fold, as to make the space globally well-defined we have to patch via a shift of the trivector.
This is a non-geometric U-duality transformation, and we can then further view the flux $\tilde f^{abc}{}_d$ as an M-theory non-geometrc $Q$-flux \cite{Blair:2014zba}.
This is the generalisation of the interpretation of non-Abelian T-dual geometries as T-folds \cite{Bugden:2019vlj}.  
 
We note that the paper \cite{Sakatani:2019zrs} considered an example where $\tilde f^{234}{}_1, \tilde f^{234}{}_2, \tilde f^{134}{}_1$, $\tilde f^{134}{}_2$ are all non-zero, in which case $S_{ab}$ has rank two (but is not diagonal in this basis), while for $\tau_{ab}$ only $\tau_{12} \neq 0$. 
For $\tilde f^{234}{}_1 = \tilde f^{234}{}_2 = 0$ this allowed other isotropic subalgebras corresponding to the embedding of the non-Abelian T-dual of the Bianchi VI algebra.
 
\subsection{Non-Abelian T-duality revisited and $CSO(3,0,2)$}

As a first example, let's consider $CSO(3,0,2)$, for which we set 
\be
S_{ab} = 4 \, \mathrm{diag}(1,1,1,0) \,,\quad \tau_{ab} = 0  \,.
\ee
We will show now how this set up actually provides an embedding for the non-Abelian T-dual (NATD) of the three-sphere $S^3$ with respect to an $SU(2)_L$ isometry sub-group.  In the M-theory section the four-dimensional geometry   with coordinates $(x^i,x^4)$, $i=1,2,3$, is given by
\be
\begin{split}
ds^2_{4} & = 
 (1+\delta_{mn}x^m x^n)^{-2/3} \left( ( \delta_{ij} + x_i x_j ) dx^i dx^j + (dx^4)^2 \right) \,,\\
C_{(3)} & = - \frac{1}{2!}\frac{\epsilon_{ijk4}  x^k}{ 1 + \delta_{mn}x^m x^n} dx^i \wedge dx^j \wedge dx^4\,.
\end{split}
\label{natdm}
\ee
With $x^4$   taken to be periodic and   identified with the M-theory circle, we can reduce to give a IIA configuration for which the 3-dimensional internal part is:
\be
\begin{split}
ds^2_{3} & =  
\frac{1}{1+\delta_{mn}x^m x^n} \left(  \delta_{ij} + x_i x_j \right) dx^i dx^j \,,\\
B_{(2)} & = - \frac{1}{2!} \frac{\epsilon_{ijk}x^k}{ 1 + \delta_{mn}x^m x^n} dx^i \wedge dx^j \,,\\
e^\Phi & = ( 1 + \delta_{mn}x^m x^n)^{-1/2} \,.
\end{split}
\label{natd}
\ee
This is indeed the aforementioned  NATD geometry.  

This prompts the obvious question:  how does the geometry prior to T-dualisation (i.e. that of the $S^3$ with round metric) manifest itself within the EDA setting?   To address this we will need to consider the EDA in the IIB-theory section.\footnote{This is natural; non-Abelian T-duality will change the chirality from IIB to IIA if three isometry generators are dualised as is the case for $SU(2)$.}     

To see this, let's look at the \EDA{} more closely.
Let's relabel our indices such that now $a=1,2,3$.   
Then the only non-zero components of the three-algebra structure constants in this case are
\be
\tilde{f}^{ab4}{}_c = -\epsilon^{ab}{}_c
\ee
where $\epsilon^{ab}{}_c \equiv \epsilon^{abd} \delta_{dc}$.

 Adapted to this we assemble the generators of the EDA as $t_a \equiv T_{a5}$, $t_4 \equiv t_{45}$,  $\tilde{t}^a \equiv \frac{1}{2} \epsilon^{abc} T_{bc}$ and $s_a = T_{a4}$ such that the algebra is given by  
\be\label{eq:semiAbelian}
[t_a, t_b ]  = 0 \, , \quad  [\tilde t^{a}, \tilde t^{b} ]  = -   \epsilon^{ab}{}_c \tilde t^{c}     \,,\quad [t_a, \tilde t^{b} ]  =  -   \epsilon^{bc}{}_a t_c \,, 
\ee
\be\label{eq:semiAbelian2}
  0 = [  t_4, t_a]=[t_4, s_a ] =[t_4, \tilde t^{b} ]\, , 
\ee
 \be\label{eq:semiAbelian3}
[t_a, s_b ]  =  +   \delta_{ab} t_4 \,,\quad 
 [s_a, s_b ]  = 0 \,,\quad
[s_a, \tilde t^{b} ]  = -2   \epsilon_{a}{}^{bc}s_c    \,, 
\ee
The original M-theory section physical subalgebra is $U(1)^4$ generated by $t_a, t_4$.  In IIA, we have a $U(1)^3$ generated by $t_a$.   In this presentation we now see an additional $SU(2)$ subalgebra  generated by $\tilde t^{a4} \equiv \frac{1}{2}\epsilon^{abc} T_{bc}$. This non-Abelian algebra is indeed a maximal isotropic in the IIB-theory section specified by the pure spinor with non-zero components $\Lambda^{45}= -\Lambda^{54}=1$.

Working now in this IIB-theory section it is easy to establish a set of generalised frame fields that realise this EDA. As detailed in the appendix, here the relevant generalised tangent bundle is $E= TM \oplus T^\star M \oplus T^\star M \oplus \Lambda^3 T^\star M$ and we use the notation $A=(a, \alpha_{(1)} ,   \tilde{\alpha}_{(1)}, \alpha_{(3)})$ to denote its sections (the generalised vectors).   
Using the type IIB generalised Lie derivative \eqref{IIBgld10}, this algebra can be realised using the following generalised frame:
\be
\begin{split} 
E^a  = \frac{1}{2} \epsilon^{abc} E_{bc} &= ( v^a , 0 , 0, 0 ) \,,\\
E_a = E_{a 5} &= (0, l_a , 0 , 0 ) \,,\\
E_{a4} & = (0,0,l_a,0) \,,\\
E_{45} &= ( 0,0,0, \mathrm{vol} ) \,,
\end{split}
\ee
where $l_a$ are the left-invariant one-forms on $\mathrm{SU}(2)$, $v^a$ the dual vector fields, and $\mathrm{vol}$ is the corresponding volume form.

Here we see that there is a natural block diagonal decomposition of the generalised frame field.  Let us consider the top left block i.e. the projections of $E^a$ and $E_{a}$ to the $O(3,3)$ generalised tangent bundle $TM \oplus T^\star M$.  These are exactly of the form of the generalised frames for Poisson-Lie duality \cite{Hassler:2017yza} in the case that the Drinfeld double is semi-Abelian of the form given in eq. \eqref{eq:semiAbelian}.  This is precisely what is required to realise non-Abelian T-duality starting with the round metric on the $S^3$.\footnote{What is used here is only an $SU(2)_L$   isometry group, so the considerations here do not directly impose   the bi-invariant metric on $S^3$.  This comes about because of the assumption made earlier in the generalised Scherk-Schwarz ansatz that $\bar{m}_{\fA \fB} = \delta_{\fA \fB}$.  Choosing other constant $\bar{m}_{\fA \fB}$ will give non-Abelian T-duals and their lifts of the $S^3$ equipped with metric $ds^2 = g^{ab} l_a \otimes l_b$ and two-form $B = b^{ab} l_{a} \wedge l_b$ with $g^{ab}$ and $b_{ab}$ constant.}  The bottom right block, i.e. the projections of $E_{a5}$ and $E_{45}$ to   $T^\star M \oplus \Lambda^3 T^\star M$ can be understood as defining a spinor representation of the $O(3,3)$ generalised frame field given by the top left block.  We shall discuss this feature in more detail when we return to the full Poisson-Lie duality context.

\subsubsection*{Relationship to Hohm-Samtleben frame}

We would like now to relate the EDA generalised frame described above to previous constructions of $\Gfour$ generalised frames realising the same $CSO(3,0,2)$ gaugings.
A particular class of generalised frames realising $CSO(p,q,r)$ gaugings were constructed by Hohm and Samtleben in \cite{Hohm:2014qga}.
For $q=0$, this frame depends on the coordinates $y^{\ui}$, where $\ui =1,\dots,p-1$, which are coordinates on an $S^{p-1}$,\footnote{Generalised frames describing sphere reductions in general have been constructed \cite{Lee:2014mla} and can be checked also to involve both a three-form and a trivector.}
 and we let $u \equiv \delta_{\ui \uj} y^{\ui} y^{\uj}$.
Then, the frame involves both a three-form and a trivector
\be
E_a = (u_a ,-\iota_{u_a} C_{(3)})  \,,\quad
E^{ab}  = ( 0 , \alpha u^a \wedge u^b ) + \lambda^{abc} E_c \,,
\label{HSFrame}
\ee
with a vielbein $u^i{}_a \equiv (1-u)^{1/2} \delta^i_a$, a function $\alpha = (1-u)^{1/6}$, and (writing the dualised forms) both a trivector and three-form, given by
\be
\lambda_a =  ( (1-u)^{-1/2} \delta_{\underline{i} \underline{k}} y^{\underline{k}} , 0 ) 
\,,\quad 
C^i =  ( (1-u)^{-1/2} y^{\underline{i}} K(u) , 0 )  \,.
\label{HSframelambdaC}
\ee
For $p=3,q=0,r=2$, $K(u)$ obeys the differential equation $2(1-u)u \partial_u K = (-2 + u ) K - 1$, and the solution is $K(u) = - 1/ u$.

For $CSO(3,0,2)$, the four-dimensional physical geometry encoded in this frame is $\mathbb{R}^2 \times S^2$ equipped with
\be
\begin{split}
ds^2_{4} & =
(dy^3)^2  + (dy^4)^2 + \left( 
\delta_{\ui \uj} + \frac{y_{\ui} y_{\uj}}{1-u} 
\right) dy^{\uj} dy^{\uj} \,,\\
C_{(3)} & = - \epsilon_{\ui \uk} y^{\uk} (1-u)^{-1/2} (1- \frac{1}{u}) dy^{\ui}  \wedge dy^3 \wedge dy^4\,.
\end{split}
\label{hsM}
\ee
Although the three-form looks rather complicated, the field strength is just $F_{(4)} = \mathrm{Vol}(S^2) \wedge dy^3 \wedge dy^4$.

Compactifying the coordinates $y^3, y^4$, this trivially reduces (on $y^4$, say) to a IIA configuration with $S^1 \times S^2$ internal space  
\be
\begin{split}
ds^2_{3} & = 
(dy^3)^2 + 
\left( 
\delta_{\ui \uj} + \frac{y_{\ui} y_{\uj}}{1-u} 
\right) dy^{\uj} dy^{\uj} \,,\\
B_{(2)} & = - \epsilon_{\ui \uk} y^{\uk} (1-u)^{-1/2} (1- \frac{1}{u} ) dy^{\ui}  \wedge dy^3 \,,
\end{split}
\label{hsIIA}
\ee
and a constant dilaton.
This can be T-dualised on $y^3$, in order to produce a solely metric configuration:
\be
\begin{split}
ds^2_{3} & = 
(d\tilde y^3 +  \frac{1}{ u} (1-u)^{+1/2}  \epsilon_{\ui \uj} y^{\uj} dy^{\ui})^2 
+\left( 
\delta_{\ui \uj} + \frac{y_{\ui} y_{\uj}}{1-u} 
\right) dy^{\uj} dy^{\uj}\,.
\end{split}
\ee
Taking our sphere coordinates to be $y^1 = \sin \theta \cos \phi$, $y^2 = \sin \theta \sin \phi$, where $\theta \in (0,\pi)$, $\phi \in (0, 2\pi)$, then $u = \sin^2 \theta$, $1-u = \cos^2 \theta$, and $dy^1 y^2 - dy^2 y^1 = - \sin^2 \theta d \phi$. 
As a result, the geometry becomes
\be
\begin{split}
ds^2_{3} & = 
(d\tilde y^3 -  \cos \theta d \phi )^2 
+ d\Omega_2^2\,.
\end{split}
\label{hopf}
\ee
This is the three-sphere $S^3$ described as a Hopf fibration.

\begin{figure}[h]
\centering

\begin{tikzpicture}

\coordinate (B) at (2,0);
\coordinate (A) at (4,0);
\coordinate (M) at (4,3);
\coordinate (HSA) at (-4,0);
\coordinate (HSM) at (-4,3);

\draw[thick,blue,<->] (B) node (Bnode) [left, draw,black, rounded corners=2pt,text width=3cm] {IIB on $S^3$ \eqref{hopf}} -- (A) node (Anode) [black,right,draw,rounded corners=2pt,text width=3cm] {IIA geometry of EDA frame \eqref{natd}} node (natdhere) [midway] {};
\draw (HSA) node (HSAnode) [left, draw, rounded corners=2pt, text width=3cm] {IIA geometry of HS frame  \eqref{hsIIA}};
\draw (HSM) node (HSMnode) [left, draw, rounded corners=2pt,text width=3cm] {M-theory geometry of HS frame  \eqref{hsM}};
\draw (M) node (Mnode) [right, draw, rounded corners=2pt,text width=3cm] {M-theory geometry of EDA frame  \eqref{natdm}};

\draw[thick,<->,blue] (Bnode) -- (HSAnode) node (here) [midway] {};

\draw[thick,->,blue] (-2,-1) node [below] {T on Hopf fibre} to [out=90,in=-90] (here);
\draw[thick,->,blue] (2,1) node [above] {NATD on $S^3$} to [out=-90,in=90] (natdhere);

\node[draw,inner sep=2mm,fit=(Bnode) (Anode) (Mnode),label=Dual within EDA, dashed] {};

\draw[red,thick,<->] (Mnode) -- (Anode) node [midway,right] {reduce/uplift};
\draw[red,thick,<->] (HSMnode) -- (HSAnode) node [midway,right] {reduce/uplift};

\draw[black,thick,<->] (HSMnode)-- (Mnode) [left] node (genU) [midway] {};
\draw [thick,->] (-1.25,4.5) node [above] {Postulated generalised U-dual} to [out=-90,in=90] (genU);

\end{tikzpicture}
\caption{Duality chains involving the NATD of $S^3$ and alternative $CSO(3,0,2)$ frames}
\label{fig:natdchain}
\end{figure}
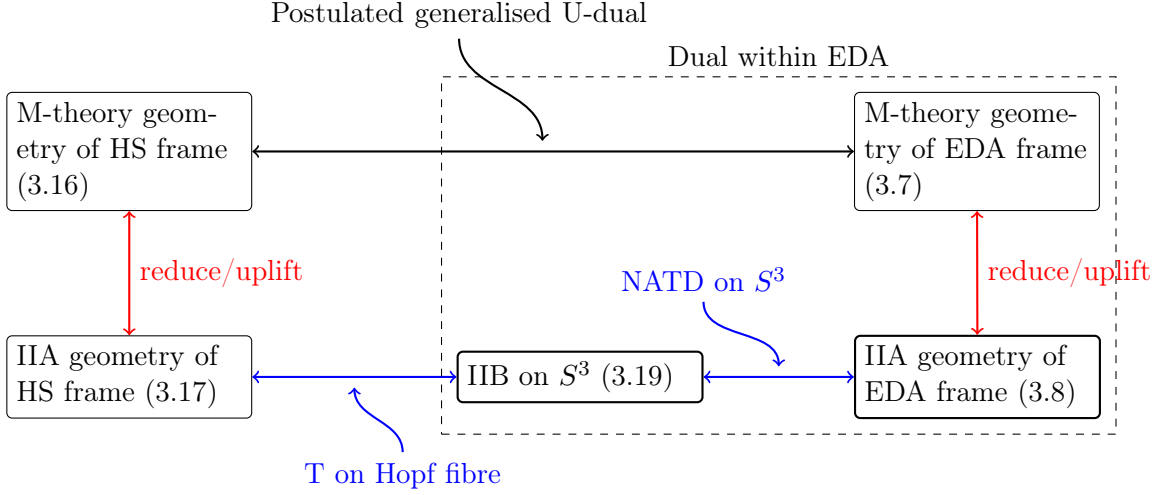

All these backgrounds produce seven-dimensional gaugings which are equivalent up to global $\Gfour$ transformations acting on the generalised fluxes.
The complete duality chain between the Hohm-Samtleben frame \eqref{HSFrame} and our EDA frame \eqref{natd} consists of: reduction from M-theory to IIA, T-duality on the Hopf fibre to IIB, non-Abelian T-duality on $S^3$ back to IIA, followed by uplift to M-theory.
This can be interpreted as a ``generalised U-duality'' however one that consists of a chain of ordinary plus non-Abelian T-dualities. Part of this duality chain takes place entirely within the EDA setting, but that involving the frame \eqref{HSFrame} uses a different construction of generalised frames.
We depict the relationships between these geometries and different $\Gfour$ frames in figure \ref{fig:natdchain}.

\subsubsection*{Non-metric 3-algebras}
A variant of the situation above is to consider the non-metric 3-algebras considered in \cite{Awata:1999dz,Gustavsson:2008dy,Gran:2008vi} for which 
\be
\tilde{f}^{ab4}{}_c = \tilde{f}^{ab}{}_c \, ,  \quad \tilde{f}^{abc}{}_d  = \tilde{f}^{ab4}{}_4 =\tilde{f}^{abc}{}_4  =0  \, , 
\ee
with $\tilde{f}^{ab}{}_c$ the structure constants of a Lie algebra.  In terms of the embedding tensor components we have equivalently 
\be
S_{44}=S_{4c}= \tau_{4c} =0 \,, \quad S_{ab} = -2 \epsilon_{cd(a} \tilde{f}^{cd}{}_{b)} \, , \quad \tau_{ab} = - \epsilon_{abc} \tilde{f}^{c d}{}_{d} \,, 
\ee
which for $\tau_{ab} =0 $ requires that $\tilde{f}$ define a uni-modular algebra.  In this case the EDA is as in \eqref{eq:semiAbelian}-\eqref{eq:semiAbelian3} after the replacement of $\epsilon^{ab}{}_c \rightarrow - \tilde{f}^{ab}{}_c$, and the construction of the IIB-theory section generalised frames goes through unchanged. This then provides an EDA embedding of non-Abelian T-duality of uni-modular group manifolds $G$ with respect to a $G_L$ isometry.  

For instance, with $S_{ab} = \mathrm{diag}(1,1,-1,0)$, such that we describe $CSO(2,1,2)$ gaugings,  we have that the non-metric three algebra is built from $SL(2)$, and that the story above will go through.  Recall that we are using $\delta_{ab}$ to contract algebra indices (i.e. not the indefinite Killing form) and hence the IIB NATD geometry above will be based on $H_3$ rather than $S^3$. 

\subsection{Euclidean 3-algebra and $CSO(4,0,1)$}

We now consider the case where $S_{ab}$ is of maximal rank:
\be
S_{ab} = 4 \, \mathrm{diag}(1,1,1,1) \,,\quad \tau_{ab} = 0 \,.
\ee
The corresponding three-algebra structure constants are totally anti-symmetric 
\be
\tilde f^{abc d} \equiv \tilde f^{abc}{}_e \delta^{ed} = \epsilon^{abcd} \,.
\label{threealgEucl}
\ee
This is well known as the unique solution of the fundamental identity for three-algebra structure constants for Euclidean three-algebras.

The four-dimensional geometry in this case is, with $x^i = (x^1,x^2,x^3,x^4)$,
\be
\begin{split}
ds^2_{4} & = 
(1+\delta_{mn} x^m x^n)^{-2/3} ( \delta_{ij} + x_i x_j ) dx^i dx^j\,,\\
C_{(3)} & = - \frac{1}{3!} \frac{1}{1+\delta_{mn} x^m x^n} \epsilon_{ijkl} x^l dx^i \wedge dx^j \wedge dx^k  \,.
\end{split}
\label{ourgeo}
\ee
The field strength is:
\be
\begin{split}
F_{(4)} & = - \frac{1}{4!} \frac{4+2\delta_{mn} x^m x^n}{(1+\delta_{mn} x^m x^n)^2}\epsilon_{ijkl} dx^i \wedge dx^j \wedge dx^k \wedge dx^l\,,\\
 & =  - (4 + 2 \delta_{mn} x^m x^n) (1+\delta_{mn} x^m x^n)^{-7/6} \mathrm{Vol}_{(4)} \,.\\
\end{split}
\ee
If we assume our coordinates are non-compact, we can write $x^i = r \hat x^i$ with $\hat x^i \hat x^j \delta_{ij} = 1$ parametrising a three-sphere, hence
\be
\begin{split}
ds^2_{4} 
 & = (1+r^2)^{1/3} \Big[ 
  dr^2 + \frac{r^2}{1+r^2} d\Omega_3^2 \Big] \,,\\
F_{(4)} 
& = - \frac{4+2r^2}{(1+r^2)^2} r^3 dr \wedge \mathrm{Vol}(S^3) \,.
\end{split}
\label{ourgeopolar}
\ee
Observe that the form of this geometry is very similar to that of the NATD geometry \eqref{natdm}, except now as seen in spherical coordinates we have an $SO(4)$ rather than $SO(3)$ isometry.

\subsubsection*{Algebra and IIB isotropics}
 
Relabelling such that $a=1,2,3$ as before, we have
\be
\tilde f^{abc}{}_4 = \epsilon^{abc} \,,\quad \tilde f^{ab4}{}_c = - \epsilon^{ab}{}_c \,.
\ee
The \EDA{} is given explicitly by the following antisymmetric brackets which indeed describe the algebra $CSO(4,0,1)$ (i.e. $\mathrm{ISO}(4)$):
\be
[t_a, t_b ]  = 0 = [t_a, t_4] \,,\quad
\ee
\be
[t_a, \tilde t^{bc} ]  =  +   \epsilon^{bc}{}_a t_4   \,,\quad
[t_a, \tilde t^{b4} ]  =  -   \epsilon^{bc}{}_a t_c \,,\quad
[t_4, \tilde t^{bc} ]  =   -   \epsilon^{bcd} t_d \,,\quad 
[t_4, \tilde t^{b4} ] =0 \,,
\ee
\be
[\tilde t^{ab}, \tilde t^{cd} ]  =  -  2    \epsilon^{cd[a} \tilde t^{b]4}\,,\quad
[\tilde t^{ab}, \tilde t^{c4} ]  = -2   \epsilon^{c[a}{}_d \tilde t^{b]d} 
 \,,\quad
[\tilde t^{a4}, \tilde t^{b4} ]  = - \epsilon^{ab}{}_c \tilde t^{c4} 
\,,
\ee
We now want to find all four- and three-dimensional subalgebras of this algebra, and check which of these are isotropic in the sense of \eqref{isotropic}.
For the Poincar\'e group in four-dimensions, the classification of all subalgebras was done in \cite{Patera:1974zd}. 
From their results we can extract that the only real isotropic subalgebras of $\mathrm{ISO}(4)$ (up to relabelling of the indices) turn out to be the four-dimensional Abelian subalgebra generated by $t_a$, along with the following three-dimensional subalgebras: $\mathrm{SU}(2)$ generated by $\tilde t^{a4}$, and $\mathrm{ISO}(2)$ generated either by $t_a, t_b, \tilde t^{c4}$ with $a\neq b \neq c$ or by $t_a, t_4$ and $\tilde t^{bc}$ with $a\neq b \neq c$.
In terms of the undualised generators, these correspond to $\{ T_{12}, T_{13}, T_{23} \}$, $\{ T_{a5}, T_{b5}, T_{ab} \} $ and $\{T_{a5} , T_{45}, T_{a4} \}$ respectively. All of these are IIB isotropics.  

Now we encounter a puzzling feature;  there are no geometric IIB uplifts of this $CSO(4,0,1)$ gauging \cite{Malek:2015hma}.
So it seems that despite the presence of a IIB isotropic we are unable to geometrically furnish this EDA within type IIB exceptional generalised geometry. This does not preclude the possibility of there being \emph{non-geometric} gaugings i.e. ones which depend on both the IIB coordinates and their duals as mentioned in \cite{Malek:2015hma}.
If this is the case, this suggests the natural home for a ``dual'' version of this frame would be in some ``deformed'' version of IIB. 
This may be analogous to, or perhaps coincide with, the so-called generalised IIB theory \cite{Arutyunov:2015mqj, Wulff:2016tju}, which necessarily arises when carrying out certain generalised T-dualities, and which can be realised in double or exceptional field theory by introducing explicit dual coordinate dependence \cite{Sakatani:2016fvh,Baguet:2016prz}, for instance see the DFT implementation of such dualities in \cite{Sakatani:2019jgu, Catal-Ozer:2019hxw}.
Although this would be interesting to develop further, we would prefer to first understand the possibility of generalised U-duality transformations between the usual 10- and 11-dimensional theories, so we leave this for future work.

\subsubsection*{Relationship to IIA on $S^3$}

Instead, let us investigate the relationship to the known $CSO(4,0,1)$ gauging arising from reduction of type IIA on $S^3$, or 11-dimensional supergravity on $\mathbb{R}\times S^3$ \cite{Cvetic:2000ah}.
Again, the idea is that any alternative frame giving rise to the same gaugings ought to provide a version of generalised U-duality.

Let us again focus on the general $CSO(p,q,r)$ frame of \cite{Hohm:2014qga}, which we wrote down in the previous subsection in \eqref{HSFrame} and \eqref{HSframelambdaC}. 
For the case $p=4$, $q=0$, $r=1$ we have coordinates $y^i = ( y^{\underline{i}}, y^z)$ where $\underline{i} = 1,2,3$, and we again define $u \equiv \delta_{\underline{i} \underline{j}} y^{\underline{i}} y^{\underline{j}}$. 
The function $K(u)$ appearing in the three-form \eqref{HSframelambdaC} is now
\be
K = - {}_2 F_1 [1, 1; 1/2;1-u] = - u^{-3/2} ( u^{1/2} + (1-u)^{1/2} \arcsin (1-u)^{1/2} )
\ee
obeying
\be
2(1-u)u \partial_u K = (-3+2u) K - 1 \,.
\label{Kprime}
\ee
This corresponds to the following four-dimensional geometry:
\be
\begin{split}
ds^2_{4} & = 
 (dy^z)^2 + \left( 
\delta_{\underline{i} \underline{j}} + \frac{y_{\underline{i}} y_{\underline{j}}}{1-u} 
\right) dy^i dy^j \,,\\
C_{(3)} & = \frac{1}{2} \epsilon_{\underline{i} \underline{j} \underline{k}} y^k (1-u)^{-1/2} (1+K(u)) dy^{\underline{i}} \wedge dy^{\underline{j}} \wedge dy^z \,,
\end{split}
\label{hsgeometry}
\ee
The coordinates $y^{\underline{i}}$ are now seen to parametrise the three-sphere $S^3$, while the isometry direction $y^z$ parametrises $\mathbb{R}$ (or $S^1$ if compact).
Thanks to the equation \eqref{Kprime} we can show that the four-form flux is constant, and this background is:
\be
\begin{split}
ds^2_{4} & = 
 (dy^z)^2 + d\Omega_3^2  \,,\\
F_{(4)} & = 2 \, \mathrm{Vol}(S^3)  \wedge dy^z \,,
\end{split}
\label{hsgeoflux} 
\ee
where $d\Omega_3^2$ is the metric on $S^3$. 
If one reduces on $y^z$, this gives IIA on $S^3$ with $H$-flux.

We therefore have two constructions of $CSO(4,0,1)$ frames.
The one based on the \EDA{} corresponds to the geometry \eqref{ourgeo}. 
This generalised frame consists of a trivial four-dimensional vielbein and a linear trivector. 
This geometry therefore has an alternative description as $\mathbb{R}^4$ (or $T^4$ if compact) carrying M-theory $Q$-flux, $Q_{a}{}^{bcd} \sim \tilde f^{bcd}{}_a$. 
The second construction is based on the geometry \eqref{hsgeoflux}, that is $\mathbb{R} \times S^3$ (or $S^1 \times {S}^3$) carrying flux of the four-form.  Unlike the case of the $CSO(3,0,2)$ gauging discussed above, there does not appear to be any easy duality chain involving conventional dualities and non-Abelian T-dualities (as in Figure \ref{fig:natdchain}) that relates the two.  Hence we believe them to be related by a novel sort of generalised U-duality transformation.

\subsection{A Leibniz geometry: $\tau_{ab} \neq 0$}
\label{Leibnizgeo} 

For an example where the EDA is not an Lie algebra, take the non-zero components of $\tau_{ab}$ to be
\be
\tau_{\alpha \beta} = \epsilon_{\alpha \beta \gamma} n^\gamma \,,\quad \alpha=1,2,3\,.
\ee
The geometry is easily seen to be
\be
\begin{split}
ds_4^2 & = \left(
1 + \frac{1}{4} (n^2 x^2 - (n\cdot x)^2 )
\right)^{-2/3}
\left( (dx^4)^2 +  \delta_{ij} dx^i dx^j + \frac{1}{4} ( \epsilon_{ijk} n^i x^j dx^k )^2 \right) \,,\\
C_{(3)} & = \frac{1}{2} \frac{1}{1 + \frac{1}{4} (n^2 x^2 - (n\cdot x)^2 )} n_i x_j dx^i \wedge dx^j \wedge dx^4\,,
\end{split}
\ee
where $n^i \equiv \delta^i_\alpha n^\alpha$, $i=1,2,3$, $n^2 \equiv \delta_{ij} n^i n^j$, $x^2 \equiv \delta_{ij} x^i x^j$, $n\cdot x \equiv \delta_{ij} n^i x^j$.
This three-form is pure gauge.

To explore the algebra, we define $u_\alpha \equiv \epsilon_{\alpha \beta \gamma} \tilde t^{\beta \gamma}$, $v^\alpha \equiv \tilde t^{\alpha 4}$, $w_\alpha \equiv t_\alpha$ and $\phi \equiv t_4$.
In this basis the M-theory section isotropic that we are considering (specified by the pure spinor $\Lambda_{\fA} = \delta_{\fA, 5} $) is the subgroup generated by $w_\alpha$ and $\phi$ with $u_\alpha$ and $v^\alpha$ the 'dual' generators.
The algebra is
\be
[u_\alpha, u_\beta] = 0 = [w_\alpha,w_\beta] = [ \phi, \mathfrak{d} ] = [ \mathfrak{d} , \phi ] \,,\quad
[v^\alpha, v^\beta] = v^{[\alpha} n^{\beta]} \,,
\ee
\be
[w_\alpha, v^\beta] = - [v^\beta, w_\alpha] = 
\frac{1}{2} ( \delta_\alpha^\beta n^\gamma w_\gamma - n^\beta w_\alpha )\,,\quad [ w_\alpha, u_\beta] = \frac{1}{2} \epsilon_{\alpha \beta \gamma} n^\gamma \phi \,,
\ee
\be
[ u_\alpha, v^\beta ] = - \frac{1}{2} ( \delta^\beta_\alpha n^\gamma u_\gamma - n^\beta u_\alpha) \,,\quad
[v^\beta, u_\alpha ] = - \frac{1}{2} ( \delta^\beta_\alpha n^\gamma u_\gamma + n^\beta u_\alpha) \,.
\ee
Notice the non-skew (i.e. Leibniz) nature of the algebra is contained entirely in the $[u,v]$ and $[v,u]$ relations, with $[u_\alpha, v^\beta ] + [ v^\beta,u_\alpha] = - \delta_\alpha^\beta n^\gamma u_\gamma$.

A second M-theory section isotropic sub-algebra is generated by $u_\alpha$ and $\phi $, which is again Abelian (this isotropic is that specified by the pure spinor $\Lambda_{\fA} = \delta_{\fA, 4})$.
 Although this simply implements interchange of the $4$ and $5$ directions,  there is no way that this new isotropic can qualify as an EDA. To see this consider the fluxes \eqref{edaflux2} which imply
 \begin{equation}
     Z_{\alpha \beta 4}{}^4 = -\frac{1}{3}\tau_{\alpha \beta} \,, \quad   Z_{\alpha \beta5}{}^5 = \frac{2}{3}\tau_{\alpha \beta} \, \, . 
 \end{equation}
To interpret this new isotropic as an EDA we must be able to find a $\tau'_{\alpha \beta}$ such that 
  \begin{equation}
     Z_{\alpha \beta4}{}^4 =  \frac{2}{3}\tau'_{\alpha \beta} \, , \quad   Z_{\alpha \beta5}{}^5 =- \frac{1}{3}\tau'_{\alpha \beta} \, , 
 \end{equation}
and there is no such $\tau'_{\alpha \beta}$.  This can be traced to the fact that the $[w,v]$ bracket is skew whilst the $[u,v]$ is not.  The fact that we can find M-theory isotropics for which the EDA conditions are not satisfied seems to point towards a possible relaxation of some of the constraints of EDA.   
 
 The sub-algebra given by $v^i$ and $\phi$ does not correspond to an M-theory section isotropic but that given by the $v^i$ alone does correspond to a IIB-theory section isotropic.

\section{Embedding Drinfeld doubles}
\label{edd}

\subsection{Decomposing the \EDA}

The embedding of Drinfeld doubles inside the exceptional Drinfeld algebra has been outlined already in \cite{Sakatani:2019zrs}. Here we expand on the discussion in that paper by systematically explaining how the Drinfeld double algebra is extended using a spinor representation, including the explicit form of the generalised frames and constraints that are needed to realise this in generalised geometry.
Then, we describe explicitly how this works for the example of the Bianchi II - Bianchi V Drinfeld double, which in \cite{Malek:2019xrf} was found to be a solution to a coboundary ansatz in the EDA.
This realises an explict example where both $f_{ab}{}^c$ and $\tilde f^{abc}{}_d$ are non-zero, and demonstrates as well one useful feature of the EDA approach which is that it geometrises the dilaton of Poisson-Lie duality.

We can describe the embedding of Drinfeld doubles by restricting to four-dimensional algebras containing a three-dimensional Lie subalgebra such that, setting $a=1,2,3$,
\be
[ T_{a5}, T_{b5} ] =   f_{ab}{}^c T_{c5} \,,\quad 
[ T_{a5}, T_{45} ] =   f_{a4}{}^4 T_{45} \,,
\label{restriction1}
\ee
and by further restricting
\be
\tilde f^{ab4}{}_c \equiv  \tilde f^{ab}{}_c  \neq 0 \,,\quad
\tilde f^{abc}{}_d = 
\tilde f^{abc}{}_4 = 0 = \tilde f^{ab4}{}_4 \,,\quad \tau_{45} = 0 \,.
\label{restriction2}
\ee
Geometrically, we assume that $v_a$ and $l^a$ obey the defining group manifold relations with the three-dimensional structure constants $f_{ab}{}^c$, while we take
\be
\lambda^{ab4} = - \pi^{ab} \,,\quad
\lambda^{abc} = 0 \,,\quad 
v_4 = \alpha \partial_4  \,,\quad l^4 = \alpha^{-1} dx^4 \,,\quad 
\ee
where we now require that $\alpha$ be a function of the three-dimensional coordinates $x^i$ such that $L_{v_a} \ln \alpha \equiv - f_{a4}{}^4$ which ensures starting with \eqref{dlambda} that $\pi^{ab}$ obeys the condition satisfied by the Poisson-Lie bivector:
\be
d\pi^{ab} = - \tilde f^{ab}{}_c l^c - 2 l^c  f_{cd}{}^{[a} \pi^{b] d}\,.
\label{ddpiab}
\ee 
Starting from \eqref{generalbg}, the above restrictions lead to the following NSNS sector geometry: 
 \be
\begin{split} 
ds^2_{10} & = ds_7^2 + \frac{1}{1+\lambda_c \lambda^c} ( \delta_{ab} + \lambda_a \lambda_b) l^a \otimes l^b   \,,\\
B_{(2)} & = \frac{1}{2} \frac{1}{1+\lambda_c \lambda^c} \epsilon_{abc} \lambda^a l^b \wedge l^c\,,\\
e^\phi & = \alpha^{-1} (1 +\lambda_c \lambda^c)^{-1/2} \,.
\end{split} 
\ee 
Extracting $G_{ab}$ and $B_{ab}$, the coefficients of the left-invariant forms, it is quick to check that  
\begin{equation}
[(G - B  )^{-1}]^{ab} = \delta^{ab}+ \pi^{ab} \, ,
\end{equation}
which is exactly the form required for a Poisson-Lie geometry \cite{Klimcik:1995dy}. (Again, we could extend this beyond the case $g_{ab} = \delta_{ab}$ by taking a more general matrix $\bar{m}_{\fA \fB}$ in \eqref{SS}.)

We now turn to the decomposition of the exceptional Drinfeld algebra \eqref{EDAf}.
We group the generators as $t_A = (t_a, t^{a4})$, $\hat t^\alpha = ( t_{4}, t^{ab} )$.
In terms of $O(3,3)$ representations, the set $t_A$ form a vector and the set $\hat t^\alpha$ form a Majorana-Weyl spinor.
The isotropy condition \eqref{isotropic} is equivalent to:
\be
\eta^{AB} t_A t_B \big|_{\mathfrak{g}}= 0\,,\quad 
\Gamma^A{}_{ \alpha\beta} t_A t^\beta \big|_{\mathfrak{g}}= 0  \,,
\ee
where $\eta_{AB}$ is the usual $O(3,3)$ metric with components $\eta_a{}^b= \eta^b{}_a = \delta^b_a$, $\eta_{ab} = \eta^{ab} = 0$,
and $\Gamma_A$ is an $O(3,3)$ gamma matrix, see appendix \ref{howtogamma}. 

After decomposing the EDA brackets \eqref{edaflux2} using \eqref{restriction1} and \eqref{restriction2} (see the explicit details in appendix \ref{howtogamma}), and regrouping into $SO(3,3)$ covariant quantities, we find the algebra is 
\be
\begin{split}
[t_A, t_B] & =  F_{AB}{}^C t_C \,,\\
[t_A, \hat t^{\alpha} ] & =  \frac{1}{4} F_{AB}{}^C (\Gamma^B{}_C)^\alpha{}_\beta \hat t^{\beta} - \frac{1}{2} \tau_{A} \hat t^{\alpha} \,,\\
[\hat t^{\alpha} , t_A ] & 
= -[t_A, \hat t^{\alpha} ] + \frac{1}{4} \left( \frac{1}{6} F_{BCD} (\Gamma_A{} \Gamma^{BCD})^\alpha{}_\beta 
- ( \Gamma_A \Gamma^B )^\alpha{}_\beta \tau_B \right)\hat t^\beta \,,\\
[\hat t^\alpha, \hat t^\beta ] & = 0 \,,
\end{split}
\label{ddspinoralg}
\ee
where the Drinfeld double structure constants $F_{AB}{}^C$, which obey $F_{ABC} \equiv F_{AB}{}^D \eta_{CD} = F_{[ABC]}$, have the expected non-zero components
\be
F_{ab}{}^c = f_{ab}{}^c \,,\quad F^{ab}{}_c = \tilde f^{ab}{}_c \,,
\ee
and we also have\footnote{This corresponds to the usual $O(d,d)$ trombone defined using the generalised dilaton $d$ via $\tau_A = E^M{}_A \partial_M ( -2 d ) + \partial_M E^M{}_A$, where $E^M{}_A$ is the $O(d,d)$ generalised vielbein (corresponding to \eqref{drinIIAvec}).
For us, $e^{-2d} = \alpha^{2} \det l$.} 
\be
\tau_a = -2 f_{a4}{}^4+f_{ac}{}^c \,,\quad \tau^a = - \tilde f^{ac}{}_c\,.
\label{tauDD}
\ee
Observe that in the second line of \eqref{ddspinoralg} we have the natural action of the Drinfeld double generators in the spinor representation. Then in the third line we have a novel action of the spinor representation on the algebra generators $t_A$, which makes this extension of the Drinfeld double into a Leibniz algebra in general.

This is not always possible due to the closure condition, as already noted in this context in \cite{Sakatani:2019zrs}, which requires
\be
f_{ab}{}^c \tilde f^{ab}{}_d = 0 \,.
\label{ddclosure}
\ee
This also follows from the general condition for a half-maximal gauging to admit an uplift to the maximal theory \cite{Dibitetto:2012rk}, see appendix \ref{trunc}.

Next, we can write down the corresponding generalised frames. 
Formally, we should decompose the exceptional tangent bundle into IIA language.
Letting $M$ denote the three-dimensional manifold, we introduce the doubled tangent bundle $\mathcal{E} \cong TM \oplus T^*M$, whose sections pair vectors and one-forms, plus a bundle $\mathcal{S} \cong \mathbb{R} \oplus \Lambda^2 T^*M$, whose sections pair functions and two-forms. 
The former bundle gives the $O(3,3)$ vector representation while the latter gives a four-dimensional spinor representation.
These appear in the decomposition $\mathbf{10} = \mathbf{6} \oplus \mathbf{4}$ of the antisymmetric representation of $\Gfour$. 

Given $V=(v,\lambda_{(1)}) \in \mathcal{E}$ and $S = (\sigma_{(0)}, \sigma_{(2)}) \in \mathcal{S}$ the generalised Lie derivative inherited from the exceptional geometry is:
\begin{align}
\mathcal{L}_V & V^\prime = ( L_{v} v^\prime, L_v \lambda_{(1)}^\prime - \iota_{v^\prime} d \lambda_{(1)} ) \in \mathcal{E} \,,
\label{VV}
\\
\mathcal{L}_V & S  = ( L_{v} \sigma_{(0)} , L_v \sigma_{(2)} + d \lambda_{(1)} \sigma_{(0)} )  \in \mathcal{S} \,,
\label{VSA}
\\
\mathcal{L}_S& V  = ( - L_v \sigma_{(0)} , -\iota_v d \sigma_{(2)} - \lambda_{(1)} \wedge d \sigma_{(0)}) \in \mathcal{S}\,,
\label{SVA} 
\end{align}
while $\mathcal{L}_S S^\prime = 0$.

We now reorganise our $\Gfour$ frame $E_{\fA \fB}$ into an $O(d,d)$-vector valued frame $E_A = (E_{a}, E^a)$, where $E^a = \frac{1}{2} \epsilon^{abc} E_{bc}$, and a spinor-valued frame, $\hat E^\alpha = ( \hat E^{0} , \hat E^{ab} )$, where $\hat E^{0} \equiv E_{45}$, $\hat E^{ab} \equiv \frac{1}{2} \epsilon^{abc} E_{c4}$. 

The vector-valued frame $E_A$ gives as sections of $TM \oplus T^*M$ 
\be
E_a = ( v_a ,  0 ) \,,\quad 
E^a = ( \pi^{ab} v_b , l^a ) \,,
\label{drinIIAvec}
\ee
which is what we expect for the Drinfeld double \cite{Hassler:2017yza}, while the spinor frame gives as sections of $\mathbb{R} \oplus \Lambda^2 T^*M$
\be
\hat E^{0} = \alpha ( 1 ,0 )
\,,\quad
\hat E^{ab}  = \alpha ( \pi^{ab} , l^a \wedge l^b ) \,.
\label{drinIIAspinor}
\ee
In the IIB case, the only change we need to make is to take the spinors to have opposite chirality, i.e. the spinor bundle now consists of odd $p$-forms, $\bar{\mathcal{S}} \cong T^*M \oplus \Lambda^3 T^*M$.
Given $S = (\sigma_{(1)}, \sigma_{(3)}) \in \bar{\mathcal{S}}$ the corresponding generalised Lie derivatives are (inherited from \eqref{IIBgld10}):
\begin{align}
\mathcal{L}_V S& = ( L_{v} \sigma_{(1)} , L_v \sigma_{(3)} - d \lambda_{(1)} \wedge \sigma_{(1)} )  \in \bar{\mathcal{S}} \,,
\label{VSB}
\\
\mathcal{L}_S V & = ( - \iota_v d \sigma_{(1)} , d \sigma_{(1)} \wedge \lambda_{(1)} ) \in \bar{\mathcal{S}}\,,
\label{SVB} 
\end{align}
and again $\mathcal{L}_S S^\prime= 0$.
The IIB spinor frame is then
\be
\hat E^{a} = \alpha ( l^a ,0 )
\,,\quad
\hat E^{abc}  = \alpha ( 3  \pi^{[ab} l^{c]}, l^a \wedge l^b \wedge l^c) \,.
\label{drinIIBspinor}
\ee
Although we can always construct the vector and spinor frames for a given Drinfeld double, they will not always obey the Leibniz algebra \eqref{ddspinoralg}.
Indeed, we have to ensure that the algebra generates constant structure constants, which leads to constraints:
\be
\pi^{[ab} \tilde f^{c]d}{}_d =0\,,\quad
f_{bc}{}^a \pi^{bc} + 2 f_{b4}{}^4 \pi^{ab} = 0 \,,
\label{ddpi}
\ee
which also follow from the constraints \eqref{adjinv} from the point of view of the \EDA.
In addition, the closure condition \eqref{ddclosure} must hold.

In this way we have also recovered a result directly from an M-theory perspective that the RR fields compatible with PL T-duality are essentially constant $O(d,d)$ spinors dressed by the spinor representation of the generalised frame field.  This was seen from a DFT perspective in  \cite{Hassler:2017yza,Demulder:2018lmj} and from a Courant algebroid approach  \cite{Severa:2018pag}.

\subsection{Example: Bianchi II and V}

\subsubsection*{Bianchi II + $\mathrm{U}(1)$ in M-theory}

This example of an \EDA{} was found in \cite{Malek:2019xrf} by requiring the three-algebra structure constants to be determined as a coboundary ansatz. 
This gives an M-theory solution where the physical subalgebra is Bianchi II + $U(1)$. 
The Bianchi II algebra, or Heisenberg algebra, can be described in a basis $\{ t_1,t_2,t_3\}$ where the single non-vanishing structure constant is $f_{23}{}^1 =1$.
The corresponding group data, including the trivial $U(1)$ factor with generator $t_4$, and $\alpha=1$, is:
\be
l^a = ( dx^1 -x^3 dx^2  , dx^2 , dx^3 , d x^4 ) \,,\quad
v_a = ( \partial_1, \partial_2 + x_3 \partial_1, \partial_3 , \partial_4 ) \,.
\ee
A trivector obeying \eqref{dlambda} is
\be
\lambda_a = ( 0, x^3 , - x^2 ,0 ) \,,
\ee
with $\tilde f^{124}{}_2 = \tilde f^{134}{}_3 = 1$.
From the above, this describes an embedding of a dual three-dimensional subalgebra with structure constants $\tilde f^{12}{}_2=\tilde f^{13}{}_3 = 1$, corresponding to the known Bianchi II / Bianchi V Drinfeld double (see \cite{Snobl:2002kq} for a classification of six dimensional doubles). 

The M-theory geometry is
\be
\begin{split}
ds^2_4  & = \frac{1}{(1+(x^2)^2 + (x^3)^2)^{2/3}}
\big(
(dx^1 - x^3 dx^2)^2 
+ (1+(x^3)^2) (dx^2)^2 
+ (1+(x^2)^2) (dx^3)^2
\\ & \qquad\qquad\qquad\qquad\qquad\qquad\qquad
- 2 x^2 x^3 dx^2 dx^3 
+ (dx^4)^2
\big)\,,\\
C_{(3)} & =  \frac{1}{1+(x^2)^2 + (x^3)^2} \left(
\frac{1}{2} d( (x^2)^2 + (x^3)^2 ) \wedge dx^1 \wedge dx^4 
+ (x^3)^2 dx^2 \wedge dx^3 \wedge dx^4 
\right)\,,
\end{split}
\ee
where $dC_{(3)} = 0$.
Reducing on the $U(1)$ direction gives a IIA geometry with
\be
\begin{split}
ds^2_3 & = \frac{1}{1+(x^2)^2 + (x^3)^2}
\big(
(dx^1 - x^3 dx^2)^2 
+ (1+(x^3)^2) (dx^2)^2 
+ (1+(x^2)^2) (dx^3)^2
\\ & \qquad\qquad\qquad\qquad\qquad
- 2 x^2 x^3 dx^2 dx^3 
\big) \,,\\
H_{(3)} & = 0 \,,\\
e^\phi &= ( 1+(x^2)^2 + (x^3)^2 )^{-1/2}\,,
\end{split}
\ee
which matches the known geometry of a Drinfeld double based on the groups Bianchi II and Bianchi V.  It is worth remarking that the physical dilaton that arises here  was implicitly constrained by the EDA.  In conventional T-duality the Buscher procedure can be used to ascertain the form of the dilaton (from the determinant produced by Gaussian elimiantion of gauge fields).  However there is no similar technique for PL duality, and determining the form of the dilaton requires either some heavy work \cite{Jurco:2017gii} or DFT techniques \cite{Demulder:2018lmj}.  The answer here was mandated by the EDA and is in agreement with these approaches.

\subsubsection*{Bianchi V in IIB}

We now have to supply the embedding of the dual Bianchi V description, in type IIB. 
Now the dual structure constants are $\tilde f^{23}{}_1 =1$ while the physical ones are $f_{12}{}^2 = f_{13}{}^3 = 1$.
A choice of group data is
\be
l^a = ( d\tilde x^1 , e^{\tilde x^1} d\tilde x^2 , e^{\tilde x^1} d\tilde x^3 ) \,,\quad
v_a = ( \partial_1 , e^{-\tilde x^1} \partial_2 , e^{-\tilde x^1} \partial_3 ) \,.
\ee
We have to pick a bivector that not only satisfies the usual Poisson-Lie condition \eqref{ddpiab} but also the conditions \eqref{ddpi} that ensure the IIB vector plus spinor frame embeds into the \EDA.
With $f_{a4}{}^4 = 0$, this requires that $\pi^{12} = \pi^{13} =0$.
Then from \eqref{ddpiab} we find that $\pi^{23}$ must obey $d\pi^{23} = (-1 + 2 \pi^{23}) l^1$, and the solution vanishing at the origin is 
\be
\pi^{23} = \frac{1}{2} ( 1-   e^{2\tilde x^1 }  ) \, .
\ee 
The corresponding physical geometry with string frame metric is
\be
\begin{split}
ds^2_3 & = 
(d\tilde x^1)^2 + 
\frac{e^{2\tilde x^1} }{1+(\pi^{23})^2} \left(
 (d\tilde x^2)^2 + (d\tilde x^3)^2
\right) 
\,,\\
B_{(2)} & = - \frac{\pi^{23} e^{2\tilde{x}^1 }}{1 + (\pi^{23})^2} d\tilde x^2 \wedge d\tilde x^3 \,,\\
e^{\phi} & =  (1 + ( \pi^{23} )^2 )^{-1/2}\,.
\end{split}
\ee

\section{Discussion}

The goal of this paper was to make geometrically concrete the algebraic structures introduced in \cite{Sakatani:2019zrs, Malek:2019xrf}.
These ``exceptional Drinfeld geometries'' provide generalised parallelisable spaces with a non-trivial relationship between the more complicated geometry and the simpler generalised frame based on a group manifold and the trivector. 
We have now developed an interesting first set of examples where the exceptional Drinfeld algebra can be explicitly connected to geometries.

A primary motivation for the introduction of the Exceptional Drinfeld Algebras was to generalise the Drinfeld double algebras that appear in generalised T-duality. 
As a confidence-building measure, we have described in detail how to embed $O(3,3)$ Drinfeld doubles and Poisson-Lie T-duality into the $\Gfour$ Drinfeld algebras. We saw that not all Drinfeld doubles can be embedded; that there are constraints that must be obeyed by their structure constants and by the explicit choice of Poisson-Lie bivector; and furthermore that the extension of the Drinfeld double requires introducing a ``spinor'' representative of the Drinfeld double and defining a non-trivial Leibniz algebra in which this acts in turn on the vector representation.

We also studied simple EDA examples where we only allowed the three-algebra structure constants to be non-zero, $\tilde f^{abc}{}_d$. These can all be realised by a simple trivector ansatz, linear in the coordinates.
In some sense, these geometries are the analogues of what should be obtained after non-Abelian T-duality, and indeed here we could reproduce the usual non-Abelian T-dual pair involving an $S^3$. 

In addition, this class of geometries can be seen to produce $CSO(p,q,r)$ gaugings of seven-dimensional maximal supergravities (with $r \geq 1$, due to the fact that at least one component of the symmetric gauging vanishes thanks to the definition of the EDA, $S_{55} = 0$).
Thus we have in effect a very simple construction of new uplifts for such gaugings.
We saw how in the $CSO(3,0,2)$ case, there was a duality chain relating our geometry to the alternative uplift due to \cite{Hohm:2014qga}, involving Hopf T-duality, non-Abelian T-duality, and M-theory uplifts.
In the $CSO(4,0,1)$ case, there appears not to be such a chain using existing notions of generalised T-dualities. 

We therefore have in this example a novel four-dimensional geometry, which encodes the Euclidean 3-algebra with $\tilde f^{abc}{}_d =\epsilon^{abc}{}_d$, and which we propose to identify as a generalised U-dual of M-theory on $\mathbb{R} \times S^3$. 
The form of this background is strikingly similar to that of the usual non-Abelian T-dual of $S^3$, suggesting that the various subtleties with the construction (for instance, how do we determine the range of the coordinates? Should we regard it as U-fold?) can be interpreted similarly as in this familiar case.

The structure of the Exceptional Drinfeld Algebra is based on the existence of isotropic subalgebras.
We had hoped to find examples in which multiple four-dimensional isotropics would be present, which could then be used as the basis for M-theory to M-theory generalised U-dualities within the EDA set-up.
Unfortunately, in the cases we have looked at, the conditions of the EDA appear to be very restrictive.
Not only does one have to have an isotropic subalgebra (and our experience shows that they are limited in number), the whole EDA is further constrained exactly such that it admits a geometric realisation in terms of just a trivector. The example of section \ref{Leibnizgeo} shows that even when there can be multiple M-theory isotropics, not all of them can be compatible with an EDA. 
Equally we saw in the $CSO(4,0,1)$ example that one can find dual IIB isotropics that do not appear to admit a geometric generalised frame description

Note that from the IIB perspective, we have not systematically reproduced the EDA from the IIB side but starting with M-theory examples considered IIB descriptions only for those cases. One therefore needs to interpret the full set of EDA structure constants in terms of a IIB construction and check whether all are geometrically realisable using a three-dimensional group manifold plus bivectors, or whether additional geometric ingredients are needed. (Similarly one might also wonder whether any information is lost in going from M-theory to IIA.)

Perhaps ultimately it may be fruitful to consider relaxing some of the axioms we used to define the EDA.
By comparison, the relaxation of the Drinfeld double (which we recall has two isotropic sub-algebras)  to having only one isotropic subalgebra is vital to describe certain models with H-flux  including the $\lambda$-deformed WZW \cite{Klimcik:2015gba}. It is likely one can also here find interesting algebras by either relaxing the group structure on $\frak{g}$ or the three-algebra structure on dual generators.

Another limitation we may have been dealing with was simple our choice of dimension.
When one goes beyond $\Gfour$ to higher-rank groups (one of us will soon report on the $\Gsix$ case \cite{EmYuhoDan}), it is likely that the number of possible constructions and transformations will be much greater. 
Other restrictions that we would hope to relax in the future would be to consider cases corresponding to less SUSY and to generalise to coset spaces rather than group manifolds.

There are also open questions related to the mathematical description of exponentiation of an EDA, when not a Lie algebra, and the precise formulation of the extended geometry in these cases.  This would likely make contact with the approach of \cite{duBosque:2017dfc} in which the physical space is identified with the quotient of an enlarged group manifold by a subgroup.
 
The algebraic structure of the exceptional Drinfeld algebra necessitated the introduction of a trivector in the generalised parallelisation.
It would be interesting to compare this with some other approaches in the literature. 
For instance, given that the idea of generalised U-duality relies on relating alternative frames giving rise to equivalent gaugings, it would be interesting to compare to the approach of  \cite{Inverso:2017lrz} which provides a systematic method for constructing frames given a set of generalised fluxes. This might also provide a method to carry out some of the generalisations mentioned above.
Further, it would be interesting to compare this construction with that of \cite{Bakhmatov:2019dow, Bakhmatov:2020kul} where the trivector is viewed as a deformation of a pre-existing geometry.

\section*{Acknowledgements}

CB, DCT and SZ all acknowledge the support of the FWO-Vlaanderen through the project G006119N and by the Vrije Universiteit Brussel through the Strategic Research Program ``High-Energy Physics''. 
CB also acknowledges the support of an FWO-Vlaanderen Postdoctoral Fellowship.
DCT is supported by The Royal Society through a University Research Fellowship {\em Generalised Dualities in String Theory and Holography} URF 150185 and in part by STFC grant ST/P00055X/1.  
We would like to thank E. Malek and Y. Sakatani for useful conversations and for detailed comments on the manuscript.

\appendix

\section{$\Gfour$ exceptional geometry}
\label{howtoexft}

\subsection{Generalised Lie derivative and generalised frames} 
\label{gld} 

Here we describe some of the technology of $\Gfour$ exceptional generalised geometry / exceptional field theory
\cite{Berman:2010is, Berman:2011cg, Coimbra:2011ky, Coimbra:2012af, Hohm:2013pua}.
We will use capital calligraphic indices $\fM,\fN,\dots = 1, \dots 5$ to label quantities transforming in the $\mathbf{5}$, and use antisymmetric pairs of such indices to label quantities transforming in the $\mathbf{10}$. 

We start with the definition of the generalised Lie derivative, which captures the bosonic local symmetries (diffeomorphisms and gauge transformations) of supergravity.
Let $A \in \mathbf{10}$ be a generalised vector of weight $\lambda_A$ and $\Lambda \in \mathbf{10}$ be a generalised vector of weight $\lambda_\Lambda = - \omega \equiv 1/5$.
The generalised Lie derivative of $V$ with respect to $\Lambda$ is
\be
\mathcal{L}_\Lambda A^{\fM \fN} 
= \frac{1}{2} \Lambda^{\fP \fQ} \partial_{\fP \fQ} A^{\fM \fN}
+ 2 \partial_{\fP \fQ} \Lambda^{\fP [\fM} A^{\fN] \fQ} 
 + \frac{1}{2} (1 + \lambda_A + \omega ) \partial_{\fP \fQ} \Lambda^{\fP \fQ} A^{\fM \fN} \,.
\label{gld10}
\ee
Meanwhile a generalised tensor $C \in \mathbf{5}$ of weight $\lambda_C$ has generalised Lie derivative
\be
\mathcal{L}_\Lambda C^{\fM} =  \frac{1}{2} \Lambda^{\fP \fQ} \partial_{\fP \fQ} C^{\fM} - C^{\fP} \partial_{\fP \fQ} \Lambda^{\fM \fQ} 
+ \frac{1}{2} ( \lambda_C + 1 + 3 \omega) \partial_{\fP \fQ} \Lambda^{\fP \fQ} C^{\fM}\,.
\label{gld5}
\ee
The actual coordinate dependence of all quantities in the theory is restricted by the formally $\Gfour$ covariant section condition
\be
\partial_{[\fM \fN} \otimes \partial_{\fK \fL]} = 0 \,,
\ee
which has independent ``solutions'' \cite{Blair:2013gqa} that break $\Gfour$ covariance and correspond to underlying M-theory, type IIA or type IIB geometries.

\subsubsection*{M-theory generalised geometry}

For the M-theory solution of the section condition, we label the $\Gfour$ indices as $\fM=(i,5)$, with $i=1,\dots,4$, and impose that $\partial_{ij} = 0$ acting on all quantities in the theory. 
Then in terms of the underlying M-theory generalised geometry we find that quantities in the $\mathbf{10}$ decompose as a pair consisting of a vector and a two-form, which are sections of (perhaps weighted) generalised tangent bundles
\be
\Lambda = ( v, \lambda_{(2)} ) \in TM \oplus \Lambda^2 T^*M \,,
\ee
\be 
A = ( a, \alpha_{(2)} ) \in ( TM \oplus \Lambda^2 T^*M ) \otimes (\det T^*M )^{( \lambda_A + \omega ) /2}\,,
\ee
and the generalised Lie derivative acts as:
\be
\mathcal{L}_\Lambda A = ( L_v a , L_v \alpha_{(2)} - \iota_a d \lambda_{(2)} ) \,,
\label{Mgld10}
\ee
where the ordinary Lie derivative $L_v$ acts on the vector $v$ and two-form $\alpha_{(2)}$ which are of weight $\lambda_A + \omega$.

Meanwhile, a generalised tensor $C$ in the fundamental corresponds to a scalar plus a three-form:
\be
C = ( c_{(0)}, c_{(3)} ) \in (\mathbb{R} \oplus \Lambda^3 T^*M ) \otimes (\det T^*M)^{(\lambda_C+3\omega)/2}\,,
\ee
and
\be
\mathcal{L}_\Lambda C = ( L_v c_{(0)} , L_v c_{(3)} + d \lambda_{(2)} c_{(0)} ) 
\label{Mgld5}
\ee
in which the ordinary Lie derivative acts on the scalar $c_{(0)}$ and three-form $c_{(3)}$ which are of weight $\lambda_C+3\omega$.

\subsubsection*{Type IIB generalised geometry}

The type IIB solution of the section condition splits $\fM = ( i , \alpha )$ with $i=1,2,3$ the spacetime index and $\alpha=4,5$ an $\mathrm{SL}(2)$ S-duality index. 
We impose $\partial_{i \alpha} =\partial_{\alpha \beta} = 0$ acting on all fields in the theory, and identify the natural derivatives with respect to the spacetime coordinates as $\partial^i \equiv \frac{1}{2} \epsilon^{ijk} \partial_k$. 
The positions of spacetime indices therefore naturally come out reversed. 

A generalised vector $A$ of weight $\lambda_A$ can now be decomposed in terms of vectors, a doublet of one-forms and a three-form:
\be
A = (a, \alpha_{(1)}, \tilde \alpha_{(1)} , \alpha_{(3)} ) \in (
TM \oplus T^*M \oplus T^*M \oplus \Lambda^3 T^*M )\otimes ( \det T^*M )^{(\lambda_A+\omega)/2} 
\ee
and with $\Lambda = ( v, \lambda_{(1)}, \tilde \lambda_{(1)}, \lambda_{(3)})$ of weight $\lambda_\Lambda = 1/5$, the generalised Lie derivative acts as
\be
\mathcal{L}_\Lambda A = ( L_v a , L_v\alpha_{(1)} - \iota_{a} d\lambda_{(1)}, L_v \tilde \alpha_{(1)} - \iota_{a} d\tilde \lambda_{(1)}, L_v \alpha_{(3)}  - d \lambda_{(1)} \wedge \tilde \alpha_{(1)}+ d \tilde \lambda_{(1)}  \wedge \alpha_{(1)}) \, ,
\label{IIBgld10}
\ee  
with the spacetime Lie derivative $L_v$ acting on the tensors here which are of spacetime weight $\lambda_A + \omega$.

A generalised tensor $C$ of weight $\lambda_C$ in the fundamental is equivalent to a one-form and a doublet of three-forms, all of spacetime weight $\lambda_C + 3\omega$:
\be
C = ( c_{(1)} , c_{(3)}, \tilde c_{(3)} )  \in  ( T^*M \oplus \Lambda^3 T^*M \oplus \Lambda^3 T^*M ) \otimes ( \det T^*M )^{( \lambda_C + 3 \omega )/ 2 }
\ee
with
\be
L_\Lambda C = (  L_v c_{(1)} ,  L_v c_{(3)} - c_{(1)} \wedge d \lambda_{(1)} , L_v \tilde c_{(3)} -c_{(1)} \wedge d \tilde \lambda_{(1)} )\,.
\label{IIBgld5}
\ee

\subsection{Generalised frames and their algebra}

The physical fields describing the geometry live in the coset $\Gfour/\Hfour$, which is parametrised by a unit determinant (inverse) generalised vielbein $\tilde E^{\fM \fN}{}_{\fA \fB} = 2 \tilde E^{[\fM}{}_{\fA} \tilde E^{\fN]}{}_{\fB}$.
The generalised vielbein $\tilde E^{\fM}{}_{\fA}$ in the $\mathbf{5}$ and that $\tilde E^{\fM \fN}{}_{\fA \fB}$ in the $\mathbf{10}$ have weight 0.
In order to construct the algebra of frame fields, we have to instead use a generalised vielbein $E^{\fM \fN}{}_{\fA\fB}$ of weight $-\omega=1/5$.
This parametrises the coset $\mathbb{R}^+ \times \Gfour / \Hfour$.
TO describe the $\mathbb{R}^+$ factor, we introduce a scalar $\Delta$ of weight $1/5$:
\be
\mathcal{L}_\Lambda \Delta = \frac{1}{2} \Lambda^{\fP \fQ} \partial_{\fP \fQ}  \Delta+\frac{1}{2} \frac{1}{5} \partial_{\fP \fQ} \Lambda^{\fP \fQ} \Delta
\ee
and define
\be
E^{\fM}{}_{\fA} = \Delta^{1/2} \tilde E^{\fM}{}_{\fA} \,\quad
E^{\fM \fN}{}_{\fA \fB} = 2 E^{[\fM}{}_{\fA} E^{\fN]}{}_{\fB} = \Delta \tilde E^{\fM \fN}{}_{\fA \fB} \,.
\ee
Hence $E^{\fM}{}_{\fA}$ is a set of 5 generalised tensors of weight $\lambda_{E_{\fA}} = 1/10$, so $\lambda_{E_{\fA}} + 3\omega = -1/2$.
Using these quantities, the algebra of generalised frames under the generalised Lie derivative can be written
\be
\mathcal{L}_{E_{\fA \fB}} E^{\fM}{}_{\fC} = - F_{\fA\fB \fC}{}^{\fD} E^{\fM}{}_{\fD} \,,
\label{5alg}
\ee
hence
\be
 \mathcal{L}_{ E_{\fA \fB}}  E^{\fM \fN}{}_{\fC \fD}
= - \frac{1}{2} F_{\fA \fB ,\,\fC \fD}{}^{\fE \fF} E^{\fM \fN}{}_{\fE \fF} 
 = 2 F_{\fA \fB[\fC}{}^{\fE} E_{\fD] E} \,,
 \label{10alg_appendix}
\ee
where
\be
F_{\fA \fB,\, \fC \fD}{}^{\fE \fF}
= 4 F_{\fA \fB [\fC}{}^{[\fE} \delta^{\fF]}_{\fD]}\,.
\ee
The form of the generalised Lie derivative means that the generalised flux $F_{\fA \fB \fC}{}^{\fD}$ can be decomposed in terms of irreducible representations of $\Gfour$
%
%
%
%
\be
F_{\fA \fB \fC}{}^{\fD} = X_{\fA \fB \fC}{}^{\fD} 
 - \frac{1}{6} \tau_{\fA \fB} \delta^{\fD}_{\fC} - \frac{1}{3} \delta^{\fD}_{[\fA} \tau_{\fB]\fC}
\label{sl5tau}
\ee
with
\be
X_{\fA \fB \fC}{}^{\fD} = Z_{\fA \fB \fC}{}^{\fD} + \frac{1}{2} \delta^{\fD}_{[\fA} S_{\fB] \fC} \,.
\ee
Here $\tau_{\fA \fB} \in \mathbf{\overline{10}}$ is the so-called trombone gauging \cite{LeDiffon:2008sh}, $S_{\fA \fB} \in \mathbf{\overline{15}}$ and $Z_{\fA \fB \fC}{}^{\fD} \in \mathbf{40}$ obeys $Z_{\fA \fB \fC}{}^{\fD} = Z_{[\fA \fB \fC]}{}^{\fD}$, $Z_{\fA \fB \fC}{}^{\fC} = 0$.
Explicit expressions in terms of the unweighted and weighted vielbeins are: 
\be
\begin{split}
\tau_{\fA \fB} & =  \Delta \left( 6 \tilde E^{\fM}{}_{\fA} \tilde E^{\fN}{}_{\fB} \partial_{\fM\fN} \ln \Delta + \partial_{\fM\fN} ( \tilde E^{\fM}{}_{\fA} \tilde E^{\fN}{}_{\fB} )\right) \\
& = 5 E^{\fM}{}_{\fA}  E^{\fN}{}_{\fB} \partial_{\fM\fN} \ln \Delta + \partial_{\fM\fN} (  E^{\fM}{}_{\fA}  E^{\fN}{}_{\fB} ) 
\end{split} 
\label{5trombone}
\ee
\be
\begin{split}
S_{\fA \fB} 
& =  4 \Delta\tilde E^{\fM}{}_{(\fA|} \partial_{\fM\fN}\tilde E^{\fN}{}_{|\fB)} 
 =  4  E^{\fM}{}_{(\fA|} \partial_{\fM\fN} E^{\fN}{}_{|\fB)} 
\end{split}
\label{tauS}
\ee
\be
\begin{split}
Z_{\fA \fB \fC}{}^{\fD} 
& 
=  \Delta \left( 3\tilde E^{\fM}{}_{[\fA}\tilde E^{\fN}{}_{\fB}\tilde E^{\fP}{}_{\fC]} \partial_{\fM\fN}\tilde E^{\fD}{}_{\fP} 
- 2 \delta^{\fD}_{[\fA|} \partial_{\fM\fN}\tilde E^{\fM}{}_{|\fB}\tilde E^\fN{}_{\fC]} \right) \\
& = 
  3
\left( E^{\fM}{}_{[\fA} E^{\fN}{}_{\fB} E^{\fP}{}_{\fC]} \partial_{\fM\fN} E^{\fD}{}_{\fP} 
-\frac{1}{2} \delta^{\fD}_{[\fA} \partial_{|\fM \fN|} ( E^{\fM}{}_{\fB} E^{\fN}{}_{\fC]} )
\right) + \frac{1}{2} \delta^{\fD}_{[\fA} \tau_{\fB \fC]}\,.
\end{split}
\label{tauZ}
\ee

\subsection{Dictionary to 11- and 10-dimensional geometries}
\label{dictionary}

The $\Gfour$ generalised geometry splits the full 11- or 10-dimensional geometry into a seven-dimensional ``external'' part and a four-dimensional ``internal'' part.
The 11- or 10-dimensional Einstein frame metric is decomposed as:
\be
ds_{11}^2 = g^{-1/5} G_{\mu\nu} dX^\mu dX^\nu + g_{ij} (dx^i + A_\mu{}^i dX^\mu ) ( dx^j + A_\nu{}^j dX^\nu ) \,,
\ee
where $G_{\mu\nu}$, $\mu,\nu = 0,\dots,6$, corresponds to a seven-dimensional Einstein frame U-duality invariant metric, and has weight $2/5$ under generalised Lie derivatives.
It is consistent to then identify
\be
\Delta = (\det G_{\mu\nu} )^{1/14} \,.
\ee
The fields carrying both external and internal indices (such as the Kaluza-Klein vector $A_\mu{}^i$) appear in the $\Gfour$ ExFT as $n$-dimensional $p$-forms in various representations of $\Gfour$. However, we will assume that these all vanish in our set-up.
We therefore have just to describe the internal metric and three-form, which together parametrise the afore-mentioned coset $\mathrm{SL}(5) / \mathrm{SO}(5)$.

\subsubsection*{M-theory parametrisation}

Start with the M-theory solution of the section condition, with physical coordinates $x^i \equiv x^{i5}$.
A conventional representation of the $\Gfour/\Hfour$ coset in terms of a (unit determinant) generalised vielbein, consistent with the diffeomorphism and gauge transformations generated by the generalised Lie derivative, is
\be
\tilde E^{\fA}{}_{\fM} = g^{1/20} \begin{pmatrix}
g^{-1/4} e^a{}_m & - g^{-1/4} e^{a}{}_n C^n \\
0 & g^{1/4} 
\end{pmatrix}\,,
\label{littleEC}
\ee
leading to a generalised metric $m_{\fM \fN} = \tilde E^{\fA}{}_{\fM} \tilde E^{\fB}{}_{\fN} \delta_{\fA \fB}$ in a five-dimensional representation
\be
m_{\fM \fN} = 
g^{1/10} 
\begin{pmatrix}
g^{-1/2} g_{mn} & - g^{-1/2} g_{mp} C^p \\
- g^{-1/2} g_{np} C^p & g^{1/2} + g^{-1/2} g_{pq} C^p C^q 
\end{pmatrix} \,,
\label{littlemC}
\ee
where the four-dimensional metric is written as $g_{mn} = e^a{}_m e^b{}_n \delta_{ab}$ and the three-form $C^m = \frac{1}{6} \epsilon^{mnpq} C_{npq}$, where $\epsilon^{1234} = 1$ is the alternating symbol.

\subsubsection*{IIB parametrisation}

The IIB solution of the section condition identifies the three-dimensional coordinates as $\tilde x_i \equiv \frac{1}{2} \epsilon_{ijk} x^{jk}$.
In this case, denote the (Einstein frame) spacetime metric by $g^{ij}$, the vielbein by $e_a{}^i$, and their determinants by $g \equiv \det (g^{ij})$, $e \equiv \det(e_a{}^i)$.
The alternating symbol in spacetime is $\epsilon^{ijk}$, and has weight $-1$, and $\epsilon_{ijk}$ has weight $+1$.
Also let $h^{\bar\alpha}{}_{\alpha}$ denote a vielbein for the coset $\mathrm{SL}(2) / \mathrm{SO}(2)$ parametrised by the axio-dilaton, with $\cH_{\alpha \beta} = h^{\bar\alpha}{}_{\alpha} h^{\bar\alpha}{}_{\alpha} \delta_{\bar \alpha \bar \beta}$.
Then the IIB geometric parametrisation takes
\be
E^{\fA}{}_{\fM} = e^{1/10} \begin{pmatrix}
e^{1/2} e_i{}^a & 0 \\
e^{-1/2} h^{\bar\alpha}{}_{\alpha} C_i^{\alpha} & e^{-1/2} h^{\bar \alpha}{}_{\alpha}
\end{pmatrix}
\,,\quad
h^{\bar\alpha}{}_{\alpha} = e^{\Phi/2} \begin{pmatrix}
1 & C_0 \\ 
0 & e^{-\Phi} 
\end{pmatrix}\,,
\label{littleECB}
\ee
\be
m_{\fM \fN} = 
g^{1/10} 
\begin{pmatrix}
g^{1/2}
g_{ij}  +  g^{-1/2} \cH_{\alpha \beta} C_i^\alpha C_j^\beta  &
 g^{-1/2}\cH_{\beta \gamma}  C_i^\gamma \\
 g^{-1/2}\cH_{\alpha \gamma} C_j^\gamma & g^{-1/2} \cH_{\alpha \beta} 
\end{pmatrix} \,,
\label{littlemCB}
\ee
with
\be
C_i^{\alpha} =\frac{1}{2} \epsilon_{ijk} ( C^{jk} , B^{jk} ) \,,\quad
\cH_{\alpha \beta} = e^\Phi \begin{pmatrix}
1 & C_0 \\
C_0 & C_0^2 + e^{-2\Phi} 
\end{pmatrix} \,.
\ee

\section{Embedding Drinfeld doubles in $\Gfour$}
\label{howtohalfmax}

\subsection{Half-maximal truncation}
\label{trunc}

In order to describe an embedding of a Drinfeld double, we can truncate the \EDA.
This means reducing from $\Gfour$ to $SO(3,3)$, along the lines of \cite{Berman:2011cg,Malek:2015hma}.
The $\mathbf{5}$ of $\Gfour$ produces one of the four-dimensional Majorana-Weyl spinor representations of (the double cover of) $SO(3,3)$ plus a singlet.
In terms of the five-dimensional indices, we write $\fM = ( I, 4)$ where $I=1,2,3,5$ is the spinorial index. 
We break $\partial_{\fM \fN} = (\partial_{I J}, \partial_{I 4})$ and impose $\partial_{I 4} = 0$.
The bispinorial derivative $\partial_{I J}$ in fact transforms in the vector representation $\mathbf{6}$ of $SO(3,3)$. 

We can compute the $O(3,3)$ generalised Lie derivative acting on the $\mathbf{5} = \mathbf{4} \oplus \mathbf{1}$, using \eqref{gld5}.
The singlet component transforms as a scalar of weight $\lambda_C + 1 + 3 \omega$ under $O(3,3)$ diffeomorphisms with parameter $\Lambda^{I J}$
\be
\mathcal{L}_\Lambda C^{4}  = 
 \frac{1}{2} \Lambda^{I J} \partial_{I J} C^{4}+ \frac{1}{2} ( \lambda_C + 1 + 3 \omega) \partial_{I J} \Lambda^{I J} C^{4}
  \,.
\ee
The spinor in the $\mathbf{4}$ transforms as:
\be
\mathcal{L}_\Lambda C^{I}  = 
 \frac{1}{2} \Lambda^{J K} \partial_{JK} C^{I}+ \frac{1}{2} ( \lambda_C + 1 + 3 \omega) \partial_{J K} \Lambda^{J K} C^{I}
  - C^{J} \partial_{J K} \Lambda^{I K} \,,
\ee
defining an $SO(3,3)$ spinorial generalised Lie derivative \cite{Berman:2011cg}.
Now, the generalised frame field $E^{\fM}{}_{\fA}$ has weight $\lambda_{E_{\fA}} = 1/10$.
Hence $E^{4}{}_{\fA}$ gives $SO(3,3)$ scalars of weight $1/2$, and $E^{I}{}_{\fA}$ gives $SO(3,3)$ spinors.
After truncating out the RR sector (by projecting out all components of the generalised vielbein carrying a single index $\fM=4$ or $\fA=4$), we are left with:
\be
E^{\fM}{}_{\fA} = \begin{pmatrix}
 E^I{}_{\alpha} & 0 \\
 0 & e^{-d} 
\end{pmatrix} \,,
\label{EdecompNoRR}
\ee
where $E^I{}_{\alpha}$ is an $SO(3,3) / SO(3) \times SO(3)$ coset element in the Majorana-Weyl spinor representation (and so has unit determinant), and $e^{-2d}$ denotes the $SO(3,3)$ generalised dilaton, which is a scalar of weight 1. 

We can now compute the algebra \eqref{5alg} of generalised frames of the form \eqref{EdecompNoRR} and interpret these in $O(3,3)$ terms.
The non-zero components of $F_{\fA \fB \fC}{}^{\fD}$ turn out to be: 
\be
F_{\alpha \beta \gamma}{}^\delta  = \tilde M_{\alpha \beta \gamma}{}^\delta + \frac{1}{2} \delta^\delta_{[\alpha} S_{\beta] \gamma}\,,\quad
F_{\alpha \beta 4}{}^4  = - \frac{1}{2} \tau_{\alpha \beta} \,,\quad
F_{\alpha 4 \beta}{}^4  = \frac{1}{2} \tau_{\alpha \beta} - \frac{1}{4} S_{\alpha \beta} \,,\\
\ee
where the irreducible fluxes have decomposed to give non-vanishing components: 
\be
\tau_{\alpha \beta} = E^{I}{}_\alpha E^J{}_{\beta} \partial_{IJ} (-2d) + \partial_{IJ} ( E^I{}_{\alpha} E^J{}_{\beta} )\,,
\quad
S_{\alpha \beta} = 4 E^I{}_{(\alpha|} \partial_{IJ} E^J{}_{|\beta)}\,,
\ee
\be
Z_{\alpha \beta \gamma}{}^\delta = \tilde M_{\alpha \beta \gamma}{}^\delta + \frac{1}{2} \delta^\delta_{[\alpha} \tau_{\beta \gamma]} \,,\quad
Z_{\alpha \beta 4}{}^4  = - \frac{1}{3} \tau_{\alpha \beta} \,,
\ee
with an $SO(3,3)$ irreducible representation 
\be
\tilde M_{\alpha \beta \gamma}{}^\delta = 3 \left( E^I{}_{[\alpha} E^J{}_{\beta} E^K{}_{\gamma]} \partial_{JK} E^\delta{}_I - \frac{1}{2} \partial_{JK} ( E^J{}_{[\alpha} E^K{}_{\beta}) \delta^\delta_{\gamma]}\right) \,,
\ee
obeying $M_{\alpha \beta \gamma}{}^\gamma = 0$.
We can more conveniently define 
\be
\tilde M^{\alpha \beta} = \frac{1}{3!} \epsilon^{\gamma\delta \epsilon \alpha} M_{\gamma\delta \epsilon}{}^\beta
 = \frac{1}{2} \epsilon^{IJKL} \partial_{I J} E^{(\alpha}{}_K E^{\beta)}{}_L 
\ee
which is symmetric.

The two irreducible symmetric representations $S_{\alpha \beta}$ and $\tilde M^{\alpha \beta}$ can be related to the self-dual and anti-self-dual parts of the usual $SO(3,3)$ generalised flux $f_{IJK}$ \cite{Dibitetto:2012rk} (using gamma matrices or equivalently 't Hooft symbols),
and a half-maximal theory uplifts to the maximal theory if \cite{Dibitetto:2012rk}
\be
S_{\alpha \beta} \tilde M^{\alpha \beta} = 0 \,.
\ee

\subsection{Drinfeld doubles} 
\label{dddetails}

So far this is a standard exercise in determining the particular fluxes of the half-maximal theory.
Now let's specialise to Drinfeld doubles. 
We break up our indices further as $I=(i,5)$ and $\alpha = (a,5)$.

\subsubsection*{Drinfeld double: IIA frame} 

To describe type IIA we take $\partial_{i 5} \neq 0$ and $\partial_{i j} = 0$.
Our data are the group manifold vector fields $v_a$, one-forms $l^a$ and the Poisson-Lie bivector $\pi^{ab}$.
We also define $\lambda_a \equiv \frac{1}{2} \epsilon_{abc} \pi^{bc}$.
Then a type IIA choice of spinorial frame and generalised dilaton is:
\be
E^I{}_\alpha =\begin{pmatrix}
(\det l)^{1/2} v^i{}_a & 0 \\
(\det l)^{-1/2} \lambda_a & (\det l)^{-1/2} 
\end{pmatrix} \,,\quad e^{-2d} = e^{-2\tilde\Phi} \det l\,.
\label{IIAframedd}
\ee
It can be checked that the following flux components are turned on:
\be
\begin{split} 
\tau_{ab} & =
\epsilon_{cd[a} \tilde f^{cd}{}_{b]}
\,,\quad
\tau_{a5}  = - 2 \partial_a \tilde \Phi + f_{ac}{}^c \,,\\
S_{ab} & = - 2 \epsilon_{cd(a} \tilde f^{cd}{}_{b)} \,,\quad S_{a5}  = - 2 f_{ac}{}^c \,,\\
\tilde M^{ab} & =  \frac{1}{2} \epsilon^{cd(a} f_{cd}{}^{b)}\,,\quad \tilde M^{a5} = \frac{1}{2} \tilde f^{ac}{}_c \,.
\end{split}
\ee
(This requires using the constraints \eqref{ddpi}, and taking the ``dilaton'' $\tilde \Phi$ to obey $\partial_a \tilde \Phi = f_{a4}{}^4$. This is not the physical dilaton but should be thought of as an extra function appearing in the definition of the frame \eqref{IIAframedd}.
To match with section \ref{edd}, take $\alpha = e^{-\tilde \Phi}$,
and in \eqref{tauDD} we have $\tau_a \equiv \tau_{a5}$ and $\tau^a \equiv \frac{1}{2} \epsilon^{abc} t_{bc}$.)

The $\Gfour$ frame in the $\mathbf{10}$ consists of a part in $\mathbf{6}$ and a part in the $\mathbf{4}$ of $SO(3,3)$.
The part in the $\mathbf{6}$ is obtained from the antisymmetrisation of the spinorial frame, $E^M{}_A \equiv 2 E^{I}{}_{[\alpha} E^J{}_{\beta]}$. 
The part in the $\mathbf{4}$ is just the spinor frame weighted by $e^{-d}$. Let's denote this by $\hat E^I{}_\alpha \equiv e^{-d} E^I{}_\alpha$.
Translating these into differential form language leads to the expressions \eqref{drinIIAvec} and \eqref{drinIIAspinor}.

\subsubsection*{Drinfeld double: IIB frame}

To describe type IIB we take: $\partial_{i 5} = 0$, $\partial_{i j} \neq 0$.
The natural partial derivatives are thus $ \partial^i = \frac{1}{2} \epsilon^{ijk} \partial_{jk}$.
Our data are now vector fields $v^a$, one-forms $l_a$ and Poisson-Lie bivector $\pi_{ab}$, with all indices in the opposite positions. 
A type IIB choice of spinorial frame and generalised dilaton is: 
\be
E^I{}_\alpha =\begin{pmatrix}
(\det l)^{-1/2} l_a{}^i  & -(\det l)^{-1/2} l_b{}^i \lambda^b \\
0 & (\det l)^{1/2} 
\end{pmatrix} \,,\quad e^{-2d} = e^{-2\tilde \Phi} \det l
\ee
where $\lambda^a = \frac{1}{2} \epsilon^{abc} \pi_{bc}$.
It can be checked that the following flux components are turned on:
\be
\begin{split} 
\tau_{ab} & =  \epsilon_{abc} ( - 2 \partial^c\tilde \Phi + f^{cd}{}_d ) \,,\quad
\tau_{a5} 
= - \tilde f_{ab}{}^b\,,\\
S_{ab} & = - 2 \epsilon_{cd(a} f^{cd}{}_{b)} \,,\quad S_{a5}  = - 2 \tilde f_{ac}{}^c \,,\\
\tilde M^{ab} & =  \frac{1}{2} \epsilon^{cd(a} \tilde f_{cd}{}^{b)} \,\quad \tilde M^{a5}  =\frac{1}{2} f^{ac}{}_c \,.
\end{split}
\ee
(Again this used the constraints \eqref{ddpi}.)

We can again translate the frame into differential form language, leading to the expressions \eqref{drinIIAvec} and \eqref{drinIIBspinor} (with indices in the opposite placement).

\subsubsection*{Uplift condition}

The condition $S_{\alpha \beta} \tilde M^{\alpha \beta} = 0$ can be easily seen to imply that a Drinfeld double uplifts to an \EDA{} only if: 
\be
 \tilde f^{ab}{}_c f_{ab}{}^c 
= 0 \,,
\ee
which is indeed the condition found in \cite{Sakatani:2019zrs} by checking closure.

\subsection{Spinors and gamma matrices}
\label{howtogamma} 

Let $e^a$ denote a vielbein basis of one-forms, and $e_a$ the inverse. 
We can represent an $O(d,d)$ spinor as a polyform, $C = \sum_p C_{(p)}$ and the gamma matrices using the wedge and interior products:
\be
\Gamma^a = \sqrt{2} e^a \wedge \,,\quad
\Gamma_a = \sqrt{2} \iota_{e_a} \,,
\label{gammageometric} 
\ee
obeying the $O(d,d)$ Clifford algebra $\{ \Gamma_a , \Gamma^b \} = 2 \delta_a^b$, $\{ \Gamma_a , \Gamma_b \} = 0$, $\{ \Gamma^a, \Gamma^b \} = 0$.

The Majorana-Weyl representations correspond to even and odd polyforms. For $d=3$, we can write these as:
\be
C_{\text{even}} = C_0 + \frac{1}{2} C_{ab} e^a \wedge e^b\,,\quad
C_{\text{odd}} =\frac{1}{6} \epsilon_{abc} (  C^0 e^a \wedge e^b \wedge e^c + 3 C^{ab} e^c ) \,,  
\ee
or in index notation $C_\alpha = (C_0, C_{ab})$, $C^\alpha = (C^0, C^{ab})$. 
Acting with a single gamma matrix maps between these representations. Acting with two gamma matrices on $C_{\text{even}}$ we obtain the antisymmetric combination $(\Gamma_{AB})_\alpha{}^\beta$ with non-zero components
\be
\begin{split} 
	(\Gamma_{ab})_0{}^{cd} & = - 4 \delta^{[c}_a \delta^{d]}_b 
\,,\quad
	(\Gamma^{ab})_{cd}{}^{0} = + 4 \delta^{[a}_c \delta^{b]}_d \,, \\
	(\Gamma_{a}{}^b)_0{}^0  & = \delta_a^b \,,\quad
	(\Gamma_{a}{}^{b})_{cd}{}^{ef} = 2 \delta_a^b \delta^{[e}_c \delta^{f]}_d + 8 \delta^b_{[c} \delta^{[e}_{d]} \delta^{f]}_{a}\,.
\end{split}
\label{gammadownup}
\ee
Similarly, acting on $C_{\text{odd}}$ we obtain the components of $(\Gamma_{AB})^\alpha{}_{\beta}$:
\be
\begin{split} 
	(\Gamma_{ab})^{cd}{}_{0} & = -  4 \delta_{[a}^c \delta_{b]}^d 
\,,\quad
	(\Gamma^{ab})^0{}_{cd}  = +  4 \delta_{[c}^a \delta_{d]}^b \,, \\
	(\Gamma_{a}{}^b)^0{}_0  & =- \delta_a^b \,,\quad
	(\Gamma_{a}{}^{b})^{cd}{}_{ef} = - 2 \delta_a^b \delta^{[c}_e \delta^{d]}_f - 8 \delta^b_{[e} \delta^{[c}_{f]} \delta^{d]}_{a}\,.
\end{split} 
\label{gammaupdown}
\ee
For convenience, let us record here also the     reduction of the EDA relations that can be encoded in the algebra \eqref{ddspinoralg} using these gamma matrices.
We have vector on vector brackets
\be
\begin{split}
[t_a,t_b] & =  f_{ab}{}^c t_c \,,\quad [t^{a4}, t^{b4} ]  =  \tilde f^{ab}{}_c t^{c4}
\\  [t_a, t^{b4} ] & =  ( - f_{ac}{}^b t^{c4} + \tilde f^{bc}{}_a t_c ) = - [t^{b4}, t_a] \,,\\
\end{split}
\ee
vector on spinor brackets 
\be
\begin{split}
[ t_a, t_4 ] &= f_{a4}{}^4 t_4 \,,\quad 
[ t_a, t^{bc} ] = ( 2 f_{ad}{}^{[b} t^{c]d} - \tilde f^{bc}{}_a t_4 + f_{a4}{}^4 t^{bc} ) \,,\\
[ t^{a4} , t_4 ]& =  \frac{1}{2} f_{bc}{}^a t^{bc} \,,\quad 
[ t^{a4}, t^{bc} ] = -2 \tilde f^{a[b}{}_d \tilde t^{c]d} \,,\\
\end{split} 
\ee
and the spinor on vector brackets
\be
\begin{split}
[  t_4,t_a ] &=  -  f_{a4}{}^4 t_4 \,,\quad
[  t^{bc},t_a ]= ( 3 f_{[de}{}^{[b} \delta^{c]}_{a]} t^{de} + \tilde f^{bc}{}_a t_4 - 3 f_{d4}{}^4 \delta^{[b}_a t^{cd]}) \,,\\
[ t_4, t^{a4}  ] &=  f_{b4}{}^4 \tilde t^{ab} \,,\quad
[  t^{bc},t^{a4} ] = - \tilde f^{bc}{}_d t^{ad}  \,,\\
\end{split} 
\ee
while the spinor on spinor brackets vanish.

\bibliography{CurrentBib}

\end{document}